\documentstyle[prd,preprint,tighten,aps,eqsecnum,%
amssymb,newlfont,epsfig]{revtex}
\newcommand{\phii}{%
{\varphi^0_i}^{\hskip+1pt \raisebox{-1pt}{$\scriptstyle\prime$}}%
}
\newcommand{\phij}{%
{\varphi^0_j}^{\hskip+1pt \raisebox{-1pt}{$\scriptstyle\prime$}}%
}
\newcommand{\phik}{%
{\varphi^0_k}^{\hskip+1pt \raisebox{-1pt}{$\scriptstyle\prime$}}%
}
\newcommand{\be}{\begin{equation}}
\newcommand{\ee}{\end{equation}}
\newcommand{\ba}{\begin{eqnarray}}
\newcommand{\ea}{\end{eqnarray}}
\newcommand{\no}{\nonumber \\}

\newcommand{\intk}{\int\! {\cal D} k\, \sum_i}
\newcommand{\ppsl}{p \hskip-5pt /}
\newcommand{\ksl}{\!\! \not \! k}
\newcommand{\psl}{\!\! \not \! p}

\newcommand{\um}{\!\! \not \! p_1}
\newcommand{\dois}{\!\! \not \! p_2}
\newcommand{\tres}{\!\! \not \! p_3}
\newcommand{\quatro}{\!\! \not \! p_4}
\begin{document}
\preprint{
\begin{tabular}{r}
UWThPh-2002-10\\
April 2002
\end{tabular}
}
\draft

\title{Soft lepton-flavor violation \\
in a multi-Higgs-doublet seesaw model}
\author{Walter Grimus\thanks{E-mail: grimus@doppler.thp.univie.ac.at}}
\address{Institut f\"ur Theoretische Physik, Universit\"at Wien \\
Boltzmanngasse 5, A--1090 Wien, Austria}
\addtocounter{footnote}{2}
\author{Lu\'{\i}s Lavoura\thanks{E-mail: balio@cfif.ist.utl.pt}}
\address{Universidade T\'ecnica de Lisboa \\
Centro de F\'\i sica das Interac\c c\~oes Fundamentais \\
Instituto Superior T\'ecnico, P--1049-001 Lisboa, Portugal}

\maketitle

\begin{abstract}
We consider the Standard Model with an arbitrary number $n_H$
of Higgs doublets and enlarge the lepton sector by adding
to each lepton family $\ell$
a right-handed neutrino singlet $\nu_{\ell R}$.
We assume that {\em all the Yukawa-coupling matrices are diagonal},
but the Majorana mass matrix $M_R$
of the right-handed neutrino singlets is an arbitrary symmetric matrix,
thereby introducing {\em an explicit but soft\/}
violation of all lepton numbers.
We investigate lepton-flavor-violating processes within this model.
We pay particular attention
to the large-$m_R$ behavior of the amplitudes for these processes,
where $m_R$ is the order of magnitude of the matrix elements of $M_R$.
While the amplitudes for processes like $\tau^- \to \mu^- \gamma$
and $Z \to \tau^+ \mu^-$ drop as $1/m_R^2$ for arbitrary $n_H$,
processes like $\tau^- \to \mu^- e^+ e^-$ and $\mu^- \to e^- e^+ e^-$
obey this power law only for $n_H = 1$.
For $n_H \geq 2$,
on the contrary,
those amplitudes do not fall off when $m_R$ increases, 
rather they converge towards constants.
This ``non-decoupling'' of the right-handed scale occurs because of
the sub-process $\ell^- \to {\ell^\prime}^- {S_b^0}^\ast$, 
where $S_b^0$ is a neutral scalar which subsequently decays to $e^+ e^-$.
That sub-process has a contribution from charged-scalar exchange which,
for $n_H \geq 2$, does
{\em not\/} decrease when $m_R$ tends to infinity.
We also perform a general study of the non-decoupling and argue that,
at the one-loop level,
after performing the limit $m_R \to \infty$
and after removing the $\nu_R$ from the Lagrangian,
our model becomes a normal multi-Higgs-doublet Standard Model
with loop-suppressed flavor-changing Yukawa couplings.
Finally,
we show that in our model
the branching ratios of all lepton-flavor-changing processes
are several orders of magnitude smaller than present experimental limits,
if one makes the usual assumptions
about the mass scales in the seesaw mechanism.
\end{abstract}


\newpage

\section{Introduction}

Recent experimental evidence strongly suggests that neutrinos mix and are,
therefore,
massive \cite{nuexperiments-atm,nuexperiments-solar,pontecorvo,reviews}.
This raises the question of why are the neutrino masses
so much smaller than the masses of charged fermions.
A simple answer to this question
is provided by the seesaw mechanism \cite{seesaw}.
In a seesaw model there are two mass scales:
\begin{description}
\item $m_D$,
the scale of the Dirac mass terms
linking the (known) left-handed neutrinos to (new) right-handed neutrinos;
\item $m_R$,
the scale of the Majorana mass terms of the right-handed neutrinos.
\end{description}
When $m_R \gg m_D$
the mixing of the two types of neutrinos gets suppressed by $m_D / m_R$
and the left-handed neutrinos acquire Majorana masses
of order $m_D^2 / m_R$,
hence much smaller than the Dirac mass terms.
Now,
$m_R \gg m_D$ is natural,
since the Majorana mass terms of the right-handed neutrinos
are gauge-invariant;
therefore,
they do not need to be of order of the Fermi scale $m_F$
of the spontaneous breaking of the electroweak symmetry.

In the context of the seesaw mechanism an interesting option
consists in having lepton-flavor symmetries
which are respected by the Dirac mass terms,
but broken by the Majorana mass terms of the right-handed neutrinos
\cite{Lbar+seesaw,GL01}.
This option arises because the Dirac mass terms
originate in Yukawa couplings of the leptons to scalar doublets,
which are hard (dimension four),
while the Majorana mass terms are soft (dimension three).
Lepton-flavor breaking thus becomes soft and,
in fact,
an unsuppressed reflection at low energies
of some physics at ultra-high energies.
This attractive hypothesis has the advantage that it allows one
to construct simple models
which explain the apparent maximal mixing of atmospheric neutrinos,
solar neutrinos,
or both simultaneously \cite{GL01}.

Once it is accepted that lepton-flavor breaking is soft,
there is nothing against introducing many scalar doublets
with Yukawa couplings to the leptons,
since lepton-flavor-changing neutral interactions
are automatically absent at tree level from those couplings.
The question then arises of knowing whether lepton-flavor-violating decays,
which arise at loop level,
are suppressed by some powers of $m_D / m_R$ or $m_F / m_R$,
or not.
Moreover,
one would like to identify the effective field theory at low scale
which corresponds to the limit $m_R \to \infty$.

In this paper we try and answer the questions above
by computing 
the lepton-flavor-violating decay $\tau^- \to \mu^- e^+ e^-$
in the context of a seesaw model with an arbitrary number of scalar doublets
and with softly-broken lepton numbers.
We take this tau decay as a concrete example for the study of 
general features of our model.
Since we compute the full one-loop decay amplitude
for $\tau^- \to \mu^- e^+ e^-$,
we simultaneously also have the amplitudes for $\tau^- \to \mu^- \gamma$,
$Z \to \tau^+ \mu^-$,
and $S^0_b \to \tau^+ \mu^-$ at our disposal.
($S^0_b$ denotes a physical neutral scalar.)
Evidently,
the changes from tau decays to muon decays can be performed trivially
in our results. 
There are in the literature a number of analogous computations 
(see Refs.~\cite{valle,koerner,illana,ilakovac,okada,cvetic} 
and citations therein),
yet our work is different for the following reasons:
\begin{enumerate}
\renewcommand{\labelenumi}{\roman{enumi}.}
\item In previous works the seesaw mechanism
is considered in the context of the Standard Model,
i.e.\
with only one Higgs doublet.
There are then no physical charged scalars,
and the neutral scalar---the Higgs particle---has been neglected
because its Yukawa couplings are suppressed by the smallness
of the charged-lepton masses.
In the present work we consider charged scalars in the loops,
and also neutral scalars $S_b^0$ in the process
\begin{equation}
\label{nzmxl}
\tau^- \to \mu^- {S_b^0}^\ast \to \mu^- e^+ e^-\, .
\end{equation}
\item In previous works all external momenta have been set to zero.
In this paper we give exact expressions for non-zero external momenta.
Since the masses of the light neutrinos are much smaller
than the masses of the charged leptons,
it does not seem justified to treat the former exactly
while neglecting the latter.
\item We study how the various contributions to the decay amplitude
behave as functions of $m_R$.
We demonstrate that the contributions previously computed
are proportional to $1/m_R^2$,
and thus negligible for sufficiently high $m_R$,
while on the other hand
some contributions to the process~(\ref{nzmxl}) remain unsuppressed.
\end{enumerate}

In order to perform our computation
in the context of a general multi-scalar-doublet model,
we had to extend the formalism previously
developed for the scalar particles in that model \cite{grimus89}.
This formalism is presented in detail in Appendix \ref{app:formalism};
it may be useful for other computations in that general model.
We also took a close look at the one-loop renormalization
of flavor-changing interactions;
we show in Appendix \ref{app:renormalization} that
the fermion wave-function renormalization,
including fermion mixing,
does not introduce any contributions to flavor-changing decays
beyond those given by diagrams with flavor-changing self-energies
in the external fermion legs \cite{soares}.

This paper is organized as follows.
In Section~\ref{sec:yukawa}
we discuss the leptonic couplings and the necessary formulas concerning
the seesaw mechanism.
Section~\ref{sec:notation} treats some notation
for the process $\tau^- \to \mu^- e^+ e^-$,
our example decay.
Section~\ref{sec:orders} deals with the orders of magnitude in our model and
Section~\ref{sec:conventions} introduces conventions and sub-processes
for the example decay.
Sections~\ref{sec:gamma},
\ref{sec:Z},
\ref{sec:S},
and \ref{sec:box} describe,
respectively, 
the photon,
$Z$,
neutral-scalar,
and box-diagram sub-processes of $\tau^- \to \mu^- e^+ e^-$.
In Section~\ref{sec:asymptotic} we discuss the limit of infinite
right-handed scale.
Section~\ref{sec:decayrates} presents decay rates
for the example decay and for other flavor-changing decays
whose amplitudes have been implicitly calculated
in Sections~\ref{sec:gamma}--\ref{sec:S}.
The conclusions are found in Section~\ref{sec:conclusions}.

\section{The leptonic couplings}
\label{sec:yukawa}

\subsection{General seesaw framework}

We consider an extension of the standard model with three families
and three right-handed neutrinos.
We label the latter with family lepton numbers:
$\nu_{eR}$,
$\nu_{\mu R}$,
and $\nu_{\tau R}$.
At the moment we do not assume conservation of the family lepton numbers,
thus this labelling has no physical content.
The Yukawa Lagrangian of the leptons is
\be
{\cal L}_\mathrm{Y} = 
- \sum_{k=1}^{n_H}\, \sum_{\ell,\ell^\prime = e, \mu, \tau}
\left[ \left( \begin{array}{cc} \varphi_k^-, & {\varphi_k^0}^\ast
\end{array} \right) \bar \ell_R \left( \Gamma_k \right)_{\ell \ell^\prime}
+ \left( \begin{array}{cc} \varphi_k^0, & - \varphi_k^+
\end{array} \right) \bar \nu_{\ell R}
\left( \Delta_k \right)_{\ell \ell^\prime} \right]
\left( \begin{array}{c} \nu_{\ell^\prime L} \\ \ell_L^\prime
\end{array} \right) + \mathrm{H.c.}
\label{Yukawa}
\ee
The mass matrix of the charged leptons $M_\ell$
and the Dirac neutrino mass matrix $M_D$ are
\be
M_\ell = \frac{1}{\sqrt{2}}\, \sum_k v_k^\ast \Gamma_k
\quad \mathrm{and} \quad
M_D = \frac{1}{\sqrt{2}}\, \sum_k v_k \Delta_k\, ,
\ee
respectively.
Without loss of generality,
we assume $M_\ell$ to be diagonal with real and positive diagonal elements:
$M_\ell = \mathrm{diag} \left( m_e, m_\mu, m_\tau \right)$.
The mass terms for the neutrinos are
\be \label{MR}
- \bar \nu_R M_D \nu_L
- \frac{1}{2}\, \bar \nu_R C M_R \bar \nu_R^T
+ \mathrm{H.c.}\, ,
\ee
where $C$ is the charge-conjugation matrix
and $M_R$ is non-singular and symmetric. 

The left- and right-handed neutrinos
are written as linear superpositions
of six physical Majorana neutrino fields $\chi_i$:
\be
\nu_{\ell L} = \sum_i \left( U_L \right)_{\ell i} \gamma_L \chi_i
\quad \mathrm{and} \quad
\nu_{\ell R} = \sum_i \left( U_R \right)_{\ell i} \gamma_R \chi_i\, ,
\ee
where $\gamma_L = \left( 1 - \gamma_5 \right) / 2$
and $\gamma_R = \left( 1 + \gamma_5 \right) / 2$
are the projectors of chirality.
The fields $\chi_i$ satisfy $\chi_i = C \bar \chi_i^T$.
The matrix
\begin{equation}
{\cal U} = \left( \begin{array}{c} U_L \\ U_R^\ast \end{array} \right)
\end{equation}
is $6 \times 6$ unitary;
therefore,
\ba
U_L U_L^\dagger &=& 1_{3 \times 3}\, , \label{uni1}
\\
U_R U_R^\dagger &=& 1_{3 \times 3}\, , \label{uni2}
\\
U_L U_R^T &=& 0_{3 \times 3}\, , \label{uni3}
\ea
and
\be
U_L^\dagger U_L + U_R^T U_R^\ast = 1_{6 \times 6}\, .
\ee
${\cal U}$ is defined in such a way that
\be
{\cal U}^T \left( \begin{array}{cc} 0 & M_D^T \\
M_D & M_R \end{array} \right) {\cal U} = \hat m
= \mathrm{diag} \left( m_1, m_2, \ldots, m_6 \right),
\label{rgets}
\ee
with real and non-negative $m_i$.
Therefore,
\ba
U_L^\ast \hat m U_L^\dagger &=& 0_{3 \times 3}\, , \label{hrtsx} \\
U_R \hat m U_L^\dagger &=& M_D\, , \\
U_R \hat m U_R^T &=& M_R\, . \label{udxmb}
\ea

The charged-current Lagrangian is
\be
{\cal L}_\mathrm{cc} = \frac{g}{\sqrt{2}}\, \sum_{\ell,i} \left[
W_\mu^- \left( U_L \right)_{\ell i}
\bar \ell \gamma^\mu \gamma_L \chi_i
+
W_\mu^+ \left( U_L^\dagger \right)_{i\ell}
\bar \chi_i \gamma^\mu \gamma_L \ell
\right].
\ee
The interaction of the $Z$ boson with the leptons is given by
\ba
{\cal L}_\mathrm{nc} & = & \frac{g}{4 c_w}\, Z_\mu
\sum_{i,j} \bar \chi_i \gamma^\mu
\left[ \gamma_L \left( U_L^\dagger U_L \right)_{ij}
- \gamma_R \left( U_L^T U_L^\ast \right)_{ij} \right]
\chi_j \label{jcksl} 
\\ & &
+ \frac{g}{c_w} Z_\mu \sum_\ell \bar \ell \gamma^\mu \left(
s_w^2 - \frac{1}{2} \gamma_L \right) \ell \,. 
\ea
When extracting the vertex from Eq.~(\ref{jcksl})
one must multiply by a factor 2,
since the neutrinos are Majorana fields.

The Yukawa couplings of the charged scalars $S_a^\pm$ to the leptons
are written in the following general notation:
\be
{\cal L}_\mathrm{Y} \left( S^\pm \right) = \sum_{a, i, \ell} \left\{
S_a^- \bar \ell \left[ \left( R_a \right)_{\ell i} \gamma_R
- \left( L_a \right)_{\ell i} \gamma_L \right] \chi_i
+ S_a^+ \bar \chi_i \left[ \left( R_a^\dagger \right)_{i\ell} \gamma_L
- \left( L_a^\dagger \right)_{i\ell} \gamma_R \right] \ell \right\}.
\ee
The notation for the scalar sector,
and the precise meaning of the $n_H$-vectors $a$,
are explained in Appendix~\ref{sub:Sab}.
One has
\ba
R_a = \Delta_a^\dagger U_R\, ,
& \mathrm{with} &
\Delta_a = \sum_k a_k \Delta_k\, , \label{La} \\
L_a = \Gamma_a U_L\, ,
& \mathrm{with} &
\Gamma_a = \sum_k a_k^\ast \Gamma_k\, . \label{Ra}
\ea
Similarly,
the Yukawa couplings of the neutral scalars $S^0_b$ 
(see again Appendix \ref{sub:Sab})
to the leptons are written
\ba
{\cal L}_\mathrm{Y} \left( S^0 \right) &=& - \frac{1}{\sqrt{2}}\,
\sum_b S^0_b \left\{
\sum_{i,j} \bar \chi_i \left[ \left( F_b \right)_{ij} \gamma_L
+ \left( F_b^\dagger \right)_{ij} \gamma_R \right] \chi_j
\right. \label{udjzg} \\ & & \left.
+ \sum_{\ell, \ell^\prime}
\bar \ell \left[ \left( \Gamma_b \right)_{\ell \ell^\prime} \gamma_L
+ \left( \Gamma_b^\dagger \right)_{\ell \ell^\prime} \gamma_R \right]
\ell^\prime \right\},
\label{neutralYuk}
\ea
with $\Gamma_b = \sum_k b_k^\ast \Gamma_k$.
The matrices $F_b$ are symmetric:
\be
F_b = \frac{1}{2}
\left( U_R^\dagger \Delta_b U_L + U_L^T \Delta_b^T U_R^\ast \right),
\quad \mathrm{with} \quad
\Delta_b = \sum_k b_k \Delta_k\, .
\label{Ab}
\ee
When extracting the Feynman rule for the vertex with the neutrinos
from Eq.~(\ref{udjzg}),
one should multiply by a factor 2,
since the $\chi_i$ are self-conjugate fields.

In the case of the charged Goldstone bosons $G^\pm = S^\pm_{a_W}$
one has (see Appendix A.3) $\left( a_W \right)_k
= g v_k / \left( 2 m_W \right)$.
Therefore,
\ba
\Delta_{a_W} &=& \frac{g}{\sqrt{2} m_W}\, M_D\, , \label{DeltaaW} \\
\Gamma_{a_W} &=& \frac{g}{\sqrt{2} m_W}\, M_\ell\, , \label{GammaaW} \\
\ea
and
\ba
R_{a_W} &=& \frac{g}{\sqrt{2} m_W}\, U_L \hat m\, , \label{RaW} \\
L_{a_W} &=& \frac{g}{\sqrt{2} m_W}\, M_\ell U_L\, . \label{LaW}
\ea

\subsection{Our model}

We assume that the Yukawa-coupling matrices $\Gamma_k$ and $\Delta_k$
are simultaneously diagonal \cite{GL01}.
Therefore,
the matrices $\Gamma_a$,
$\Delta_a$,
$\Gamma_b$,
and $\Delta_b$ are all diagonal.
The neutrino Dirac mass matrix $M_D$ is also diagonal,
hence
\be
U_L \hat m U_R^\dagger = M_D^\dagger
\quad \mathrm{and} \quad
U_L \hat m^2 U_L^\dagger = M_D^\dagger M_D
\quad \mathrm{are\ diagonal}.
\label{our}
\ee
Now the labelling of the neutrino fields according to family lepton
numbers acquires a well-defined meaning.

Diagonal Yukawa-coupling matrices are achieved
by assuming invariance of the Yukawa Lagrangian
under $U(1)_{L_\alpha}$ ($\alpha = e,\mu,\tau$),
the groups associated with conservation of the lepton number $L_\alpha$
for each lepton family.
Since the gauge part of the Standard-Model Lagrangian
is invariant under these $U(1)$ symmetries anyway,
and since the scalar doublets do not transform under these $U(1)$ groups,
the only place where these lepton-number symmetries are violated
is the Majorana mass term of the right-handed singlets in Eq.~(\ref{MR}).
Since the mass term is an operator of dimension three,
this violation is soft \cite{GL01}.
Hence,
the one-loop amplitudes for lepton-flavor-violating processes
must be finite. 

\section{Notation for the process}
\label{sec:notation}

The flavor-changing decay that we want to study is
\be
\tau^- \left( p_1 \right) \rightarrow
\mu^- \left( p_2 \right) e^+ \left( p_3 \right) e^- \left( p_4 \right).
\label{process}
\ee
Clearly,
\ba
p_1^2 &=& m_\tau^2\, , \\
p_2^2 &=& m_\mu^2\, , \\
p_3^2 = p_4^2 &=& m_e^2\, .
\ea

We denote
\be
\begin{array}{rcccl}
q^\alpha &=& p_1^\alpha - p_2^\alpha &=& p_3^\alpha + p_4^\alpha\, ,
\\
r^\alpha &=& p_1^\alpha - p_4^\alpha &=& p_2^\alpha + p_3^\alpha\, ,
\end{array}
\ee
and
\be
p_+^\alpha = \frac{p_1^\alpha + p_2^\alpha}{2}\, .
\ee
Then,
\be
\begin{array}{rcccl}
4 m_e^2 &\le& q^2 &\le& \left( m_\tau - m_\mu \right)^2\, ,
\\
\left( m_\mu + m_e \right)^2 &\le& r^2 &\le& \left( m_\tau - m_e \right)^2\, .
\end{array}
 \label{isjxl}
\ee

The amplitude $M$ for the process of Eq.~(\ref{process})
involves the Dirac spinors $\bar u_\mu = \bar u_\mu \left( p_2 \right)$,
$u_\tau = u_\tau \left( p_1 \right)$,
$\bar u_e = \bar u_e \left( p_4 \right)$,
and $v_e = v_e \left( p_3 \right)$.
These spinors satisfy
\ba
\bar u_\mu \dois &=& m_\mu \bar u_\mu\, , \\
\um u_\tau &=& m_\tau u_\tau\, , \\
\bar u_e \quatro &=& m_e \bar u_e\, , \label{ue} \\
\tres v_e &=& - m_e v_e\, . \label{ve}
\ea

In our calculations we shall often need the the following product
of elements of $U_L$:
\be\label{sumxi}
x_i = \left( U_L \right)_{\mu i} \left( U_L^\dagger \right)_{i\tau}\, ,
\quad \mathrm{satisfying} \quad
\sum_i x_i = 0 \,,
\ee
where the second relation
is a consequence of unitarity---see Eq.~(\ref{uni1}).
Furthermore,
we shall make use of dimensional regularization,
evaluating integrals in a space--time of dimension $d$.
Therefore,
we define
\be
d = 4 - 2 \epsilon
\quad \mathrm{and} \quad
{\cal D} k = d^d k / (2\pi)^d \,,
\ee
where the latter expression is an abbreviation for 
integration over the momentum $k$.
Eventually,
we shall take the limit $\epsilon \to 0$.
Then,
in some integrals the divergent constant
\be\label{K}
\mathcal{K} = \epsilon^{-1} - \gamma + \ln (4 \pi)
\ee
will appear ($\gamma$ is Euler's constant). 
The reason why $\mathcal{K}$ always drops out
when calculating the amplitude for the process~(\ref{process})
in our model will be discussed in detail.

\section{Orders of magnitude}
\label{sec:orders}

We assume that the matrix elements of $M_D$ are of order $m_D$
and the square roots of the eigenvalues of $M_R^\dagger M_R$
are of order $m_R$,
with $m_D \ll m_R$.
Then,
\be
\begin{array}{lclcl}
{\displaystyle m_i \sim \frac{m_D^2}{m_R}}
&\mathrm{and}&
{\displaystyle \left( U_R \right)_{\ell i} \sim \frac{m_D}{m_R}}
&\mathrm{for}&
i = 1,2,3\, ;
\\*[4mm]
m_i \sim m_R
&\mathrm{and}&
{\displaystyle \left( U_L \right)_{\ell i} \sim \frac{m_D}{m_R}}
&\mathrm{for}&
i = 4,5,6\, .
\end{array}
\label{Vsmall}
\ee
There is also the order of magnitude,
which we may call $m_\ell$,
of the charged-lepton masses.
This may be taken as either $m_\tau \sim 2$ GeV,
or $m_\mu \sim 10^{-1}$ GeV,
or $m_e \sim 5 \times 10^{-4}$ GeV.
Because of Eqs.~(\ref{isjxl}),
$q^2 \sim r^2 \sim m_\ell^2$.
For simplicity we shall identify $m_\ell$ with $m_D$.

Finally,
there is the Fermi scale $m_F$,
in between 10 GeV and 1 TeV.
The $W$ mass $m_W$,
the charged-scalar masses $m_a$,
and the neutral-scalar masses $m_b$ are all taken to be of order $m_F$.
The scalar-potential couplings $C_{a a^\prime b}$
in Eq.~(\ref{jxhsl}) are also assumed to be of order $m_F$.
The overall hierarchy of mass scales is thus $m_D \ll m_F \ll m_R$.

When we have information about the masses of 
the light neutrinos we can estimate the order of magnitude of $m_R$ 
via the seesaw relation \cite{seesaw} in the first line of Eq.~(\ref{Vsmall}). 
If we take the light-neutrino masses to be of the order of
$\sqrt{\Delta m^2_\mathrm{atm}} \sim 0.05$ eV,
where  $\Delta m^2_\mathrm{atm}$
is the neutrino mass-squared difference
relevant for atmospheric-neutrino oscillations \cite{nuexperiments-atm},
and if we assume that $m_D \sim m_\mu$ or $m_\tau$,
then we obtain $m_R \sim 10^8 \div 10^{11}$ GeV.
Thus we might regard $m_R \sim 10^{10}$ GeV
as a typical order of magnitude of the right-handed scale,
keeping in mind however that $m_R$ might deviate
several orders of magnitude from this value.

It is important to notice that,
with the convention $m_\ell = m_D$,
the matrices $R_{a_W}$ and $L_{a_W}$ in Eqs.~(\ref{RaW}) and (\ref{LaW}),
giving the Yukawa couplings of the charged Goldstone boson,
are of the same orders of magnitude
as the corresponding general matrices $R_a$ and $L_a$
in Eqs.~(\ref{Ra}) and (\ref{La}),
provided we make the assumption
\begin{equation}\label{orderY}
\Gamma_j \,, \Delta_j \sim m_D/m_F \,;
\end{equation}
this is a natural relation in view of $m_\ell = m_D$.
This means that the factors $m_D^2 / m_R^2$
suppressing some charged-Goldstone-boson contributions
are exactly the same as those suppressing the corresponding,
and more general,
charged-scalar contributions.

In the result for the momentum integrals large logarithms,
like for instance $\ln (m_R/m_F)$,
arise.
We shall not take into account such logarithms in our estimates
of orders of magnitude.

\section{Conventions and sub-processes}
\label{sec:conventions}

We shall compute the process~(\ref{process})
using the conventions and vertices given in Ref.~\cite{book}.
In the $Z^0$--$G^0$ sector we use the unitary gauge,
thus discarding $G^0$.
On the contrary,
in the $W^\pm$--$G^\pm$ sector we use Feynman gauge.
This means that the propagator of $W^\pm$
is $-i g_{\rho \sigma} / (k^2 - m_W^2)$.
The charged-Goldstone-boson contributions
will usually be taken into account
together with the general charged-scalar contributions;
the sums over the charged scalars $S_a^\pm$ will {\em not\/} exclude
the charged Goldstone boson $G^\pm = S_{a_W}^\pm$.
We remind that,
in Feynman gauge,
$m_{a_W} = m_W$.

The process~(\ref{process}) may proceed via box diagrams or
through one of the three following sub-processes.
In the sub-process with amplitude $\bar M_A$ the initial lepton $\tau^-$
decays into the final lepton $\mu^-$
together with a {\em virtual\/} photon $A$ with momentum $q$;
the photon later decays into the $e^+ e^-$ pair in the final state.
The corresponding amplitude may be written
\be\label{barMA}
\bar M_A = \frac{M_A^\rho}{q^2}\, e\, \bar u_e \gamma_\rho v_e\, ,
\ee
where $e$ is the positron charge and 
$M_A^\rho$ is the amplitude for $\tau^- \left( p_1 \right)
\to \mu^- \left( p_2 \right) \gamma \left( q \right)$,
the $\tau^-$ and the $\mu^-$ being on mass shell
while the photon is off mass shell.
Current conservation,
i.e.\ Eqs.~(\ref{ue}) and (\ref{ve}),
implies that one may discard all terms proportional to $q^\rho$
from $M_A^\rho$.
The sub-process with amplitude $\bar M_Z$ is analogous to the previous one
with the virtual photon substituted by a virtual $Z$ boson.
Its amplitude may be written
\be\label{barMZ}
\bar M_Z = \frac{M_Z^\rho}{q^2 - m_Z^2}
\left( g_{\rho \theta} - \frac{q_\rho q_\theta}{m_Z^2} \right)
\frac{g}{c_w}\,
\bar u_e \gamma^\theta \left( s_w^2 - \frac{\gamma_L}{2} \right) v_e\, .
\ee
Finally, 
there is the sub-process with amplitude $\bar M_b$,
in which the $\tau^-$ decays into $\mu^-$
together with an off-shell neutral scalar $S^0_b$ with momentum $q$.
The corresponding amplitude is
\be\label{barMb}
\bar M_b = \frac{M_b}{\sqrt{2} \left( q^2 - m_b^2 \right)}\,
\bar u_e \left[ \left( \Gamma_b \right)_{ee} \gamma_L
+ \left( \Gamma_b^\dagger \right)_{ee} \gamma_R \right] v_e\, .
\ee
One must sum $\bar M_b$ over all physical neutral scalars $b \neq b_Z$.

Besides these three sub-processes,
there are also box diagrams,
which are all finite,
to be considered in Section~\ref{sec:box}.

No one-loop diagram with a neutral scalar $S_b^0$ in the loop
can contribute to the process (\ref{process}).

The infinities in the amplitude for the process~(\ref{process}) cancel
for the following reasons:
\begin{enumerate}
\renewcommand{\labelenumi}{\Alph{enumi}.}
\item Conservation of the electromagnetic current;
\item unitarity of the diagonalization matrix $\mathcal{U}$;
\item flavor-diagonal Yukawa-coupling matrices.
\end{enumerate}
Item A is independent both of our model and of the seesaw mechanism,
and applies to the photon sub-process.
Concerning item B,
the relations~(\ref{uni1}) and (\ref{uni3}) are relevant,
the first one in the form of Eq.~(\ref{sumxi}).
Only item C is directly connected with our model and,
clearly,
it plays a role only in charged-scalar exchange.
More generally,
items B and C are responsible for the cancellation
of all terms independent of the neutrino masses $m_i$.

\section{$\tau^- \to \mu^- A^\ast$}
\label{sec:gamma}

\subsection{Computation}

Let us work out $M_A^\rho$ in detail.
First consider the transition $\tau^- (p) \to \mu^- (p)$,
effected either by $W^\pm$ exchange or by charged-scalar exchange.
Call the corresponding amplitude $- i \Sigma (p)$.
Then,
there are the following two contributions to $M_A^\rho$:
\ba
M_{A1}^\rho &=& - e \bar u_\mu \gamma^\rho\,
\frac{\ \um + m_\mu}{m_\tau^2 - m_\mu^2}
\left[ - i \Sigma \left( p_1 \right) \right] u_\tau\, ,
\label{A1}
\\
M_{A2}^\rho &=& - e \bar u_\mu \left[ - i \Sigma \left( p_2 \right) \right]
\frac{\ \dois + m_\tau}{m_\mu^2 - m_\tau^2}\,
\gamma^\rho u_\tau\, .
\label{A2}
\ea
We obtain for $\Sigma (p)$
\be
- i \Sigma (p) = \intk \frac{1}{D_i} \left[ \frac{g^2 x_i}{2}\,
\frac{\left( d - 2 \right) \ksl \gamma_L}{\left( k - p \right)^2 - m_W^2}
+ \sum_a \frac{X_{ai}}{\left( k - p \right) - m_a^2} \right],
\label{Sigma}
\ee
where
\ba
X_{ai} &=&
\left( R_a \right)_{\mu i}
\left( R_a^\dagger \right)_{i \tau}\,
\ksl \gamma_L
+ \left( L_a \right)_{\mu i}
\left( L_a^\dagger \right)_{i \tau}\,
\ksl \gamma_R
\no & &
- m_i \left( L_a \right)_{\mu i}
\left( R_a^\dagger \right)_{i \tau}
\gamma_L
- m_i \left( R_a \right)_{\mu i}
\left( L_a^\dagger \right)_{i \tau}
\gamma_R\, .
\ea
The first term in the right-hand-side of Eq.~(\ref{Sigma})
is the $W^\pm$-exchange contribution.
The second term automatically includes,
in the sum over $a$,
the contribution of the charged Goldstone boson---which is obtained
by setting $m_{a_W} = m_W$ and by using Eqs.~(\ref{RaW}) and (\ref{LaW})
for the matrices $R_{a_W}$ and $L_{a_W}$.

There is one more contribution to $M_A^\rho$,
namely from diagrams in which the photon attaches either to the $W^\pm$
or to the charged scalar in the loop.
There is no $W^\pm S_a^\mp \gamma$ vertex
except in the case $S_a^\mp = G^\mp$;
also---see Appendix \ref{sub:feynman}---the vertex 
$W^\pm G^\mp \gamma$ has a factor $m_W$
which cancels the denominator $m_W^{-1}$ in the Yukawa coupling of $G^\mp$.
One obtains the amplitude
\ba
M_{A3}^\rho &=& e \bar u_\mu \intk
\frac{1}{D_i} \left\{ \frac{g^2 x_i}{D_{1W} D_{2W}}\,
\left[ \left( d - 2 \right) \left( k - p_+ \right)^\rho
\ksl \gamma_L
- 4 k \cdot p_+ \gamma^\rho \gamma_L + 4 p_+^\rho \ksl \gamma_L
\right. \right. \no & & \left. \left.
+ m_\tau \left( \gamma^\rho \ksl - \ksl \gamma^\rho \right) \gamma_R
+ m_\mu \left(\, \ksl \gamma^\rho - \gamma^\rho \ksl \right) \gamma_L
\right]
+ \sum_a \frac{2 \left( k - p_+ \right)^\rho X_{ai}}
{D_{1a} D_{2a}}
\right\} u_\tau\, .
\label{A3}
\ea
We have used the shorthands
\be
D_{\iota W} = \left( k - p_\iota \right)^2 - m_W^2
\quad \mathrm{and} \quad
D_{\iota a} = \left( k - p_\iota \right)^2 - m_a^2\, ,
\quad \mathrm{for} \quad
\iota = 1, 2\, .
\ee
The sum over $a$ in Eq.~(\ref{A3}) includes
the contribution of the vertex $G^+ G^- \gamma$.
The other terms in the right-hand-side of that equation
are the contributions from the vertex $W^+ W^- \gamma$
and from the vertices $W^\pm G^\mp \gamma$.

Thus,
$M_A^\rho = M_{A1}^\rho + M_{A2}^\rho + M_{A3}^\rho$.

One writes the Feynman integrals symbolically:
\be
\int\! {\cal D} k\, \frac{1}{D_i D_{1W}} = a_1^i \, ,
\quad
\int\! {\cal D} k\, \frac{k^\theta}{D_i D_{1W}} = 
b_1^i p_1^\theta\, ;
\label{a1}
\ee
\be
\int\! {\cal D} k\, \frac{1}{D_i D_{2W}} = a_2^i \, ,
\quad
\int\! {\cal D} k\, \frac{k^\theta}{D_i D_{2W}} = 
b_2^i p_2^\theta\, ;
\ee
\be
\begin{array}{l}
{\displaystyle
\int\! {\cal D} k\, \frac{1}{D_i D_{1W} D_{2W}} = a_3^i \, ,
\quad
\int\! {\cal D} k\, \frac{k^\theta}{D_i D_{1W} D_{2W}} =
c_1^i p_1^\theta + c_2^i p_2^\theta\, ,}
\\*[5mm]
{\displaystyle
\int\! {\cal D} k\, \frac{k^\theta k^\psi}{D_i D_{1W} D_{2W}} =
d_1^i p_1^\theta p_1^\psi + d_2^i p_2^\theta p_2^\psi
+ f^i \left( p_1^\theta p_2^\psi + p_2^\theta p_1^\psi \right)
+ u^i g^{\theta \psi}.
}
\end{array}
\label{x}
\ee
In $a_1^i,\ldots, u^i$, the index $i$ indicates the dependence on $m_i^2$.
The coefficients $u^i$,
$a_1^i$,
$a_2^i$,
$b_1^i$,
and $b_2^i$ diverge when $d \to 4$,
while all other coefficients are finite.
However,
the following relations hold:
\ba
a_1^i - a_2^i &=& q^2 \left( c_1^i - c_2^i \right)
+ \left( m_\tau^2 - m_\mu^2 \right) k_3^i\, ,
\label{a1a2}
\\
b_1^i - b_2^i &=& q^2 \left( d_1^i - d_2^i \right)
+ \left( m_\tau^2 - m_\mu^2 \right) \left( k_1^i + k_2^i \right),
\\
m_\tau^2 b_1^i - m_\mu^2 b_2^i &=& 
q^2 \left( m_\tau^2 d_1^i - m_\mu^2 d_2^i \right)
+ \left( m_\tau^2 - m_\mu^2 \right) \left( 2 u^i - q^2 f^i
+ m_\tau^2 k_1^i + m_\mu^2 k_2^i \right),
\no & &
\label{b1b2}
\ea
where
\ba
k_1^i &=& d_1^i + f^i - c_1^i\, ,
\\
k_2^i &=& d_2^i + f^i - c_2^i\, ,
\\
k_3^i &=& c_1^i + c_2^i - a_3^i\, . \label{k3}
\ea
One may write relations similar to Eqs.~(\ref{a1})--(\ref{k3})
for the integrals
which have $D_{\iota a}$ instead of $D_{\iota W}$ (for $\iota = 1, 2$),
then we use the notation $a_1^{i,a}, \ldots, u^{i,a}$ in order to
indicate the dependences on $m_i^2$ and on $m_a^2$.

Using Eqs.~(\ref{a1a2})--(\ref{b1b2}) one finds that $M^\rho_A$
is finite and that,
moreover,
it respects gauge invariance,
i.e.\
$q_\rho M^\rho_A = 0$:
\ba
M^\rho_A &=& e \sum_i \bar u_\mu \left\{
i\sigma^{\rho \lambda} q_\lambda
\left( \alpha_{Li}\, \gamma_L + \alpha_{Ri}\, \gamma_R \right) \right.
\no & &
+ \beta_{Li} \left[ q^\rho \gamma_L - \frac{q^2}{m_\tau^2 - m_\mu^2}\,
\gamma^\rho \left( m_\mu \gamma_L + m_\tau \gamma_R \right) \right]
\no & & \left.
+ \beta_{Ri} \left[ q^\rho \gamma_R - \frac{q^2}{m_\tau^2 - m_\mu^2}\,
\gamma^\rho \left( m_\mu \gamma_R + m_\tau \gamma_L \right)
\right] \right\} u_\tau\, .
\label{MAgauge}
\ea
One obtains (for $d=4$)
\ba
\alpha_{Li} &=&
g^2 x_i m_\mu \left( k_2^i + c_1^i + c_2^i \right)
\no & &
+ \sum_a \left[
m_\mu \left( R_a \right)_{\mu i}
\left( R_a^\dagger \right)_{i\tau} k_2^{i,a}
+ m_\tau \left( L_a \right)_{\mu i}
\left( L_a^\dagger \right)_{i\tau} k_1^{i,a}
- m_i \left( L_a \right)_{\mu i}
\left( R_a^\dagger \right)_{i\tau} k_3^{i,a}
\right], \nonumber \\ 
&& \label{alphaLi}
\\
\alpha_{Ri} &=&
g^2 x_i m_\tau \left( k_1^i + c_1^i + c_2^i \right)
\no & &
+ \sum_a \left[
m_\tau \left( R_a \right)_{\mu i}
\left( R_a^\dagger \right)_{i\tau} k_1^{i,a}
+ m_\mu \left( L_a \right)_{\mu i}
\left( L_a^\dagger \right)_{i\tau} k_2^{i,a}
- m_i \left( R_a \right)_{\mu i}
\left( L_a^\dagger \right)_{i\tau} k_3^{i,a}
\right], \nonumber  \\ 
&& \label{alphaRi}
\\
\beta_{Li} &=&
g^2 x_i m_\mu \left( k_5^i + c_1^i + c_2^i \right)
\no & &
+ \sum_a \left[
m_\mu \left( R_a \right)_{\mu i}
\left( R_a^\dagger \right)_{i\tau} k_5^{i,a}
- m_\tau \left( L_a \right)_{\mu i}
\left( L_a^\dagger \right)_{i\tau} k_4^{i,a}
+ m_i \left( L_a \right)_{\mu i}
\left( R_a^\dagger \right)_{i\tau} \Delta c^{i,a}
\right], \nonumber \\ && \label{betaLi}
\\
\beta_{Ri} &=&
- g^2 x_i m_\tau \left( k_4^i + c_1^i + c_2^i \right)
\no & &
+ \sum_a \left[
- m_\tau \left( R_a \right)_{\mu i}
\left( R_a^\dagger \right)_{i\tau} k_4^{i,a}
+ m_\mu \left( L_a \right)_{\mu i}
\left( L_a^\dagger \right)_{i\tau} k_5^{i,a}
+ m_i \left( R_a \right)_{\mu i}
\left( L_a^\dagger \right)_{i\tau} \Delta c^{i,a}
\right], \nonumber \\
&& \label{betaRi}
\ea
where
\ba
k_4^i &=& f^i - d_1^i\, , \\
k_5^i &=& f^i - d_2^i\, , \label{k5}
\ea
and similarly for the coefficients $k_4^{i,a}$ and $k_5^{i,a}$.
Also,
$\Delta c^{i,a} = c_2^{i,a} - c_1^{i,a}$.

\subsection{Order of magnitude} \label{mxlsr}

Integrating over the momentum $k$ one finds,
from Eqs.~(\ref{x}),
\be
\left( a_3^i, c_1^i, c_2^i, d_1^i, d_2^i, f^i \right) =
 \frac{- i}{16 \pi^2}\, \int_0^1 \! dx \int_0^{1-x} \! dy\,
\frac{1}{\Delta_i^W} \left( 1, y, x, y^2, x^2, xy \right),
\ee
where
\be
\Delta_i^W = \left( x + y \right) m_W^2
+ \left( 1 - x - y \right) \left( m_i^2 - x m_\mu^2 - y m_\tau^2 \right)
- x y q^2\, .
\ee

Let us denote by $g^i$ a generic coefficient $a_3^i$,
$c_{1,2}^i$,
$d_{1,2}^i$,
or $f^i$,
and similarly $g^{i,a}$ for $a_3^{i,a}$, etc.
Clearly,
$g^i$ is a function of $m_i^2$,
with $g^i \sim m_F^{-2}$ for $i=1,2,3$
and $g^i \sim m_R^{-2}$ for $i=4,5,6$.
Moreover,
for $i=1,2,3$,
$g^i$ is independent of $m_i$
up to corrections of order
$m_i^2 / m_W^2 \sim m_D^4 / \left( m_R^2 m_F^2 \right)$.
The same holds for the $g^{i,a}$,
of course.

Considering Eqs.~(\ref{MAgauge})--(\ref{betaRi}) in detail,
we see that $M_A^\rho$ involves terms of five types:
\begin{enumerate}
\item $g^2 m_\ell^2 \sum_i x_i g^i$;
\item $m_\ell^2 \sum_i \left( R_a \right)_{\mu i}
\left( R_a^\dagger \right)_{i\tau} g^{i,a}$;
\item $m_\ell^2 \sum_i \left( L_a \right)_{\mu i}
\left( L_a^\dagger \right)_{i\tau} g^{i,a}$;
\item $m_\ell \sum_i m_i \left( R_a \right)_{\mu i}
\left( L_a^\dagger \right)_{i\tau} g^{i,a}$;
\item $m_\ell \sum_i m_i \left( L_a \right)_{\mu i}
\left( R_a^\dagger \right)_{i\tau} g^{i,a}$.
\end{enumerate}
We remind the reader that we set $m_\ell \sim m_D$.
Also,
from Eqs.~(\ref{La}) and (\ref{Ra}),
we see that $L_a$ is $U_L$ and $R_a$ is $U_R$
except for diagonal Yukawa-coupling matrices.

Terms of types 1 and 3 are similar.
They are proportional to $(U_L)_{\mu i} (U_L^\dagger)_{i\tau}$,
which is of order 1 for $i=1,2,3$
and of order $m_D^2 / m_R^2$ for $i=4,5,6$.
As we have seen,
$g^i \sim m_F^{-2}$ is almost independent of $i$ for $i=1,2,3$.
One may therefore use
\be
\sum_{i=1}^3 x_i = - \sum_{i=4}^6 x_i \sim \frac{m_D^2}{m_R^2}
\label{equilibrium}
\ee
to estimate $m_\ell^2 \sum_{i=1}^3 x_i g^i$ to be of order
$m_D^4 / \left( m_R^2 m_F^2 \right)$.
For $i=4,5,6$ one has $g^i \sim m_R^{-2}$
and $x_i \sim m_D^2 / m_R^2$,
hence terms of type 1 or 3 are of order $m_D^4 / m_R^4$.

Terms of type 2 are proportional to
$(U_R)_{\mu i} (U_R^\dagger)_{i\tau}$.
Therefore,
for $i=1,2,3$ they are of order $m_D^4 / \left( m_R^2 m_F^2 \right)$
and for $i=4,5,6$ they are of order $m_D^2 / m_R^2$.

Finally,
terms of types 4 and 5 always include a suppression $m_D/m_R$
from the mixing matrices $U_L$ and $U_R$.
Taking into account also $m_\ell$,
$m_i$,
and $g^{i,a}$,
those terms are of order $m_D^4 / \left( m_R^2 m_F^2 \right)$
for $i=1,2,3$,
or $m_D^2 / m_R^2$ for $i=4,5,6$.

In summary,
we find that the amplitude for the vertex $\tau^- \to \mu^- A^\ast$
is suppressed by $m_D^2 / m_R^2$.
With $m_D \sim 1$ GeV and $m_R \sim 10^{10}$ GeV
one has a suppression factor $10^{-20}$ in the amplitude
of $\tau \to \mu \gamma$.
Note that terms of type 2--5
contain a product of two Yukawa coupling constants. 
If we are more specific
and assume the relation (\ref{orderY}) for the Yukawa couplings,
then there is a suppression factor of order
$m_D^4/(m_R^2 m_F^2)$ for
\emph{all of them\/} and,
in addition,
a $(16\pi^2)^{-1}$ from the loop integration.
Terms of type 1,
on the other hand,
have an extra factor $g^2$.

\section{$\tau^- \to \mu^- Z^\ast$}
\label{sec:Z}

\subsection{Graphs in which the $Z$ attaches to charged particles}

The vertex $\tau^- \to \mu^- Z^\ast$
has three contributions $M_{Z1}^\rho$,
$M_{Z2}^\rho$,
and $M_{Z3}^\rho$ analogous to the corresponding contributions
to $\tau^- \to \mu^- A^\ast$.
Comparing them one easily concludes that
\be
M_{Z1}^\rho + M_{Z2}^\rho + M_{Z3}^\rho
= \frac{g \left( s_w^2 - c_w^2 \right)}{2 e c_w}\, M_A^\rho
+ \frac{g}{2 c_w}\, T^\rho\, ,
\ee
with
\ba
T^\rho &=&
- \bar u_\mu \left\{ \gamma^\rho \gamma_R\,
\frac{\ \um + m_\mu}{m_\tau^2 - m_\mu^2}
\left[ - i \Sigma \left( p_1 \right) \right]
+ \left[ - i \Sigma \left( p_2 \right) \right]
\frac{\ \dois + m_\tau}{m_\mu^2 - m_\tau^2}\,
\gamma^\rho \gamma_R \right\} u_\tau
\no & &
+ \bar u_\mu \intk \frac{g^2 x_i}{D_i D_{1W} D_{2W}} \left[
\left( 2 - d \right) \left( k - p_+ \right)^\rho \ksl \gamma_L
+ 4 k \cdot p_+ \gamma^\rho \gamma_L
- 4 p_+^\rho \ksl \gamma_L
\right. \no & & \left.
+ m_\tau \ksl \gamma^\rho \gamma_R
+ m_\mu \gamma^\rho \ksl \gamma_L
- 2 m_i^2 \gamma^\rho \gamma_L \right] u_\tau\, .
\ea
We compute $T^\rho$ by using the same method
as in Eqs.~(\ref{a1})--(\ref{k5}),
arriving at the result
\ba
T^\rho &=&
- \left( \bar u_\mu \gamma_R u_\tau \right)
\sum_i g^2 x_i m_\tau \left[ 2 p_+^\rho  \left( k_1^i + c_1^i \right)
- q^\rho \left( k_4^i + c_1^i \right) \right]
\no & & -
\left( \bar u_\mu \gamma_L u_\tau \right)
\sum_i g^2 x_i m_\mu \left[ 2 p_+^\rho  \left( k_2^i + c_2^i \right)
+ q^\rho \left( k_5^i + c_2^i \right) \right]
\no & & + \bar u_\mu \gamma^\rho \sum_i \left( L_i \gamma_L
+ R_i \gamma_R \right) u_\tau\, ,
\ea
where
\ba
L_i &=& g^2 x_i \left[ \left( 2 - d \right) u^i
- 2 m_i^2 a_3^i + \left( 2 m_\tau^2 + m_\mu^2 - q^2 \right) c_1^i
+ \left( m_\tau^2 + 2 m_\mu^2 - q^2 \right) c_2^i \right],
\\
R_i &=& g^2 x_i m_\mu m_\tau \left[ c_1^i + c_2^i - k_1^i - k_2^i
+ \frac{q^2}{m_\tau^2 - m_\mu^2} \left( d_2^i - d_1^i \right) \right]
\no & &
+ \sum_a \left\{
\left( R_a \right)_{\mu i} \left( R_a^\dagger \right)_{i\tau}
m_\mu m_\tau \left[ - k_1^{i,a} - k_2^{i,a}
+ \frac{q^2}{m_\tau^2 - m_\mu^2} \left( d_2^{i,a} - d_1^{i,a} \right) \right]
\right.
\no & & +
\left( L_a \right)_{\mu i} \left( L_a^\dagger \right)_{i\tau} \!
\left[ - 2 u^{i,a} + q^2 f^{i,a} - m_\tau^2 k_1^{i,a} - m_\mu^2 k_2^{i,a}
+ \frac{q^2}{m_\tau^2 - m_\mu^2}
\left( m_\mu^2 d_2^{i,a} - m_\tau^2 d_1^{i,a} \right) \!\right]
\no & & \left. +
m_i \left[
m_\tau \left( L_a \right)_{\mu i} \left( R_a^\dagger \right)_{i\tau}
+ m_\mu \left( R_a \right)_{\mu i} \left( L_a^\dagger \right)_{i\tau}
\right] \left[ k_3^{i,a}
+ \frac{q^2}{m_\tau^2 - m_\mu^2}
\left( c_1^{i,a} - c_2^{i,a} \right) \right] \right\}.
\ea

Let us consider the order of magnitude of $T^\rho$.
We realize at a glance that most of its terms
are of one of the five types already considered in Subsection~\ref{mxlsr}.
In $L_i$ there is a term $x_i m_i^2 a_3^i$,
which is of order $m_D^2 / m_R^2$ for $i=4,5,6$
and additionally suppressed for $i=1,2,3$.
The main originality of the $T^\rho$ is,
however,
the presence of a term with $u^i$ in $L_i$ and with $u^{i,a}$ in $R_i$.
Computing the divergent coefficient $u^i$ one finds
\be
u^i = \frac{i}{32 \pi^2}\, \int_0^1 \! dx \int_0^{1-x} \! dy\,
\left( \mathcal{K} - \ln{\Delta_i^W} \right),
\label{uskzl}
\ee
where the infinite constant $\mathcal{K}$ is defined in Eq.~(\ref{K}).
Now,
$\mathcal{K}$ is independent of $i$
and cancels out when one sums over $i$.
This happens in the case of $u^i$ because $\sum_i x_i = 0$, 
and in the case of $u^{i,a}$ because 
$(L_a L_a^\dagger)_{\mu\tau} = (\Gamma_a \Gamma_a^\dagger)_{\mu\tau} = 0$;
the last relation holds because the matrices $\Gamma_a$ are diagonal,
which is a specific property of our model.
In $u^i$, Eq.~(\ref{uskzl}), and also in $u^{i,a}$ one may,
therefore, discard the divergence.
We may also substitute $\ln{\Delta_i^W}$
by the logarithm of the dimensionless quantity
$\Delta_i^W / \Delta_1^W$.\footnote{There is
an infrared divergence when $m_i = 0$.
Therefore,
it is better to take $m_i = m_1$ as the subtraction point for the logarithm.}
This logarithm is large for $i=4,5,6$
while for $i=1,2,3$ it is of order
$m_D^4 / \left( m_R^2 m_F^2 \right)$.
The order of magnitude of $\sum_i x_i u^i$
and $\sum_i (L_a)_{\mu i} (L_a^\dagger)_{i\tau} u^{i,a}$ is,
therefore,
$m_D^2 / m_R^2$.\footnote{We remind the reader that
we do not take into account the enhancements introduced by large logarithms.}

In summary,
$\sum_{\iota =1}^3 M_{Z\iota}^\rho$
is suppressed by $m_D^2 / m_R^2$,
even in the presence of charged scalars.

\subsection{Graphs in which the $Z$ attaches to neutrinos}

There are also contributions to $M_Z^\rho$
in which the $Z$ boson attaches to the neutrino line,
thereby changing the neutrino eigenstate from $\chi_j$ to $\chi_i$.
The corresponding vertex is given by Eq.~(\ref{jcksl}),
with an extra factor 2 because of the self-conjugated character
of the neutrinos.
One obtains a contribution $M^\rho_{Z4}$ to $M_Z^\rho$,
and $M_Z^\rho = \sum_{\iota =1}^4 M_{Z\iota}^\rho$.
Let us define
\be
D_{ij}^W = x m_i^2 + y m_j^2
+ \left( 1 - x - y \right) \left( m_W^2 - x m_\mu^2 - y m_\tau^2 \right)
- x y q^2\, ,
\label{DijW}
\ee
and similarly $D_{ij}^a$,
which is identical to $D_{ij}^W$ with $m_W$ substituted by $m_a$.
We then find
\ba
M^\rho_{Z4} &=&
\frac{- i g^3}{32 \pi^2 c_w}\, \sum_{i,j}
\left( U_L \right)_{\mu i} \left( U_L^\dagger \right)_{j\tau} \,
\int_0^1 \! dx \, \int_0^{1-x} \! dy\
\bar u_\mu \left[ \frac{\xi_1^\rho \left( U_L^T U_L^\ast \right)_{ij}
+ \xi_2^\rho \left( U_L^\dagger U_L \right)_{ij}}{D^W_{ij}}
\right. \no & & \left.
+ \left( U_L^\dagger U_L \right)_{ij} \gamma^\rho \gamma_L
\left( \ln{D_{ij}^W} - \mathcal{K} + 2 \right) \right] u_\tau
\no & & 
+ \frac{- i g}{32 \pi^2 c_w}\, \sum_{i,j} \sum_a
\int_0^1 \! dx \, \int_0^{1-x} \! dy\
\bar u_\mu \left\{
\frac{ \xi_3^\rho}{D_{ij}^a}
+ \gamma^\rho \left[
\left( L_a \right)_{\mu i}
\left( U_L^\dagger U_L \right)_{ij}
\left( L_a^\dagger \right)_{j\tau} \gamma_R
\right. \right. \no & & \left. \left.
-
\left( R_a \right)_{\mu i}
\left( U_L^T U_L^\ast \right)_{ij}
\left( R_a^\dagger \right)_{j\tau} \gamma_L
\right] \left( \ln{D^a_{ij}} - \mathcal{K} + 1 \right) \right\} u_\tau\, ,
\label{ustdr}
\ea
with
\be
\xi_1^\rho = m_i m_j \gamma^\rho \gamma_L\, ,
\ee
\ba
\xi_2^\rho &=& \left[
2 p_1 \cdot p_2 \left( 1 - x \right) \left( 1 - y \right)
+ m_\mu^2 x \left( x - 1 \right)
+ m_\tau^2 y \left( y - 1 \right)
\right] \gamma^\rho \gamma_L
\no & &
+ m_\mu m_\tau \left( 1 - x - y \right) \gamma^\rho \gamma_R
\no & &
+ 2 m_\mu \left( 1 - x \right) \left[
p_2^\rho x - p_1^\rho \left( 1 - y \right) \right] \gamma_L
+ 2 m_\tau \left( 1 - y \right) \left[
p_1^\rho y - p_2^\rho \left( 1 - x \right) \right] \gamma_R\, ,
\ea
\ba
\xi_3^\rho &=&
m_i m_j \gamma^\rho \left[
\left( L_a \right)_{\mu i}
\left( U_L^T U_L^\ast \right)_{ij}
\left( L_a^\dagger \right)_{j\tau}
\gamma_R
-
\left( R_a \right)_{\mu i}
\left( U_L^\dagger U_L \right)_{ij}
\left( R_a^\dagger \right)_{j\tau}
\gamma_L
\right]
\no & & +
m_i m_\tau \left( 1 - y \right) \gamma^\rho \left[
\left( R_a \right)_{\mu i}
\left( U_L^\dagger U_L \right)_{ij}
\left( L_a^\dagger \right)_{j\tau}
\gamma_L
-
\left( L_a \right)_{\mu i}
\left( U_L^T U_L^\ast \right)_{ij}
\left( R_a^\dagger \right)_{j\tau}
\gamma_R
\right]
\no & & +
m_i x \left( m_\mu \gamma^\rho - 2 p_2^\rho \right) \left[
\left( R_a \right)_{\mu i}
\left( U_L^\dagger U_L \right)_{ij}
\left( L_a^\dagger \right)_{j\tau}
\gamma_R
-
\left( L_a \right)_{\mu i}
\left( U_L^T U_L^\ast \right)_{ij}
\left( R_a^\dagger \right)_{j\tau}
\gamma_L
\right]
\no & & +
m_j m_\mu \left( 1 - x \right) \gamma^\rho \left[
\left( L_a \right)_{\mu i}
\left( U_L^\dagger U_L \right)_{ij}
\left( R_a^\dagger \right)_{j\tau}
\gamma_L
-
\left( R_a \right)_{\mu i}
\left( U_L^T U_L^\ast \right)_{ij}
\left( L_a^\dagger \right)_{j\tau}
\gamma_R
\right]
\no & & +
m_j y \left[
\left( L_a \right)_{\mu i}
\left( U_L^\dagger U_L \right)_{ij}
\left( R_a^\dagger \right)_{j\tau}
\gamma_L
-
\left( R_a \right)_{\mu i}
\left( U_L^T U_L^\ast \right)_{ij}
\left( L_a^\dagger \right)_{j\tau}
\gamma_R
\right] \left( m_\tau \gamma^\rho - 2 p_1^\rho \right)
\no & & +
m_\mu m_\tau \left( 1 - x - y \right) \gamma^\rho \left[
\left( R_a \right)_{\mu i}
\left( U_L^T U_L^\ast \right)_{ij}
\left( R_a^\dagger \right)_{j\tau}
\gamma_R
-
\left( L_a \right)_{\mu i}
\left( U_L^\dagger U_L \right)_{ij}
\left( L_a^\dagger \right)_{j\tau}
\gamma_L
\right]
\no & & +
m_\mu x \left( 1 - x \right)
\no & & \times
\left( m_\mu \gamma^\rho - 2 p_2^\rho \right)
\left[
\left( R_a \right)_{\mu i}
\left( U_L^T U_L^\ast \right)_{ij}
\left( R_a^\dagger \right)_{j\tau}
\gamma_L
-
\left( L_a \right)_{\mu i}
\left( U_L^\dagger U_L \right)_{ij}
\left( L_a^\dagger \right)_{j\tau}
\gamma_R
\right]
\no & & +
m_\tau y \left( 1 - y \right)
\no & & \times
\left[
\left( R_a \right)_{\mu i}
\left( U_L^T U_L^\ast \right)_{ij}
\left( R_a^\dagger \right)_{j\tau}
\gamma_R
-
\left( L_a \right)_{\mu i}
\left( U_L^\dagger U_L \right)_{ij}
\left( L_a^\dagger \right)_{j\tau}
\gamma_L
\right] \left( m_\tau \gamma^\rho - 2 p_1^\rho \right)
\no & & +
x y \left[
\left( R_a \right)_{\mu i}
\left( U_L^T U_L^\ast \right)_{ij}
\left( R_a^\dagger \right)_{j\tau}
\left( 2 m_\mu p_1^\rho \gamma_L + 2 m_\tau p_2^\rho \gamma_R
- 2 p_1 \cdot p_2 \gamma^\rho \gamma_L \right)
\right. \no & & \left.
-
\left( L_a \right)_{\mu i}
\left( U_L^\dagger U_L \right)_{ij}
\left( L_a^\dagger \right)_{j\tau}
\left( 2 m_\mu p_1^\rho \gamma_R + 2 m_\tau p_2^\rho \gamma_L
- 2 p_1 \cdot p_2 \gamma^\rho \gamma_R \right)
\right].
\ea

As before,
$\mathcal{K}$ does not depend on $m_i$ and on $m_j$
and cancels out when one sums Eq.~(\ref{ustdr}) over $i$ and over $j$.
The cancellations occurs in the second line
as a consequence of Eq.~(\ref{uni1}) and in the fourth line 
as a consequence of (\ref{uni3}) and the form of $R_a$ in Eq.~(\ref{La}),
which are properties of the general seesaw framework;
in the case of the term
$(L_a)_{\mu i} (U_L^\dagger U_L)_{ij} (L_a^\dagger)_{j\tau}$,
cancellation upon summation over $i$ and over $j$
also hinges upon the fact that the matrices $\Gamma_a$ are diagonal,
which is a property of our specific model.

We are therefore free to subtract from $\ln{D^W_{ij}}$
in Eq.~(\ref{ustdr}) its value when $i=j=1$,
i.e.\ its value when both $\chi_i$ and $\chi_j$ are the lightest neutrino.
We obtain $\ln{( D^W_{ij} / D^W_{11} )}$,
which is zero for $i=j=1$,
of order $m_D^4 / \left( m_R^2 m_F^2 \right)$
when $i \in \left\{ 1,2,3 \right\}$ and $j \in \left\{ 2,3 \right\}$
or vice-versa,
and large otherwise.
We similarly substitute $\ln{D^a_{ij}}$
by $\ln{( D^a_{ij} / D^a_{11} )}$.

It is now possible to evaluate the order of magnitude
of each contribution to $M_{Z4}^\rho$.
After tedious yet straightforward consideration,
we conclude that all terms are suppressed by,
at least,
$m_D^2 / m_R^2$.
In most cases this suppression applies term by term;
in some exceptional cases one must sum the contributions
over the light neutrinos and apply Eq.~(\ref{equilibrium}).
This happens,
for instance,
with
\be
\left( U_L \right)_{\mu i} \left( U_L^\dagger U_L \right)_{ij}
\left( U_L^\dagger \right)_{j\tau} \int_0^1 \! dx \, \int_0^{1-x} \! dy \
\frac{1}{D^W_{ij}}\, ,
\ee
which is of order $m_F^{-2}$ for $i,j \in \left\{ 1,2,3 \right\}$
but acquires a suppression $m_D^2 / m_R^2$
when $i$ and $j$ are both summed over the light neutrinos.

In conclusion,
the contributions to the decay~(\ref{process})
from both photon or $Z$ exchange are all suppressed
by $m_D^2 / m_R^2$,
even in the presence of extra scalar doublets.
This suppression also ensures that the simpler processes
$\tau^- \to \mu^- \gamma$ and $Z \to \tau^+ \mu^-$
are invisible in all feasible experiments.
We next look to the contributions to the decay~(\ref{process})
from neutral-scalar exchange.

\section{$\tau^- \to \mu^- {S^0_{\lowercase{b}}}^*$}
\label{sec:S}

\subsection{Self-energy graphs}

Similarly to what happens with the couplings to the photon and $Z$ boson,
there are two self-energy graphs
for the coupling to the neutral scalar $S^0_b$:
\ba
M_{b1} &=& \frac{1}{\sqrt{2}}\, \bar u_\mu
\left( \Gamma_b \gamma_L + \Gamma_b^\dagger \gamma_R \right)_{\mu\mu}
\frac{ \not \! p_1 + m_\mu}{m_\tau^2 - m_\mu^2}
\left[ - i \Sigma (p_1) \right] u_\tau\, , \label{G3a}
\\
M_{b2} &=& \frac{1}{\sqrt{2}}\, \bar u_\mu
\left[ -i \Sigma (p_2) \right]  
\frac{ \not \! p_2 + m_\tau}{m_\mu^2 - m_\tau^2}
\left( \Gamma_b \gamma_L + \Gamma_b^\dagger \gamma_R \right)_{\tau\tau}
u_\tau\, . \label{G4a}
\ea
The difference relative to the case of the gauge bosons
lies in the fact that $S^0_b$ couples differently
to the charged leptons $\tau$ and $\mu$---in the first case
with the Yukawa coupling $\left( \Gamma_b \right)_{\tau \tau}$,
in the second case with $\left( \Gamma_b \right)_{\mu \mu}$.
As a consequence,
it is not possible to use Eqs.~(\ref{a1a2})--(\ref{b1b2})
and we must compute $\Sigma (p)$ explicitly.
Define
\ba
D_i^W &=& x m_i^2 + \left( 1-x \right) \left( m_W^2 - x p^2 \right),
\\
D_i^a &=& x m_i^2 + \left( 1-x \right) \left( m_a^2 - x p^2 \right).
\ea
The relevant integrals are then
\ba
f_i^a &=& \int_0^1 \! dx \, \ln{\frac{D_i^a}{D_1^a}}\, ,
\\
g_i^a &=& \int_0^1 \! dx \left( 1-x \right) \ln{\frac{D_i^a}{D_1^a}}\, ,
\ea
together with $f_i^W$ and $g_i^W$ analogously defined.
One has
\ba
- i \Sigma (p) &=&
\frac{-i}{16 \pi^2}\, \sum_i\, \left\{
g^2 x_i\, g_i^W\! \psl \gamma_L
+ \sum_a g_i^a \psl \left[
\left( R_a \right)_{\mu i} \left( R_a^\dagger \right)_{i\tau} \gamma_L
+ \left( L_a \right)_{\mu i} \left( L_a^\dagger \right)_{i\tau} \gamma_R
\right]
\right. \no & & \left.
- \sum_a f_i^a m_i \left[
\left( L_a \right)_{\mu i} \left( R_a^\dagger \right)_{i\tau} \gamma_L
+ \left( R_a \right)_{\mu i} \left( L_a^\dagger \right)_{i\tau} \gamma_R
\right] \right\}.
\label{sigma-exact}
\ea
As in the previous section,
in the original definitions of $f_i^a$ and $g_i^a$
there should be a divergence $\mathcal{K}$ added to the logarithms.
However,
$\mathcal{K}$ is $i$-independent and yields a null contribution
to $\Sigma (p)$ upon summation over $i$.
This happens because of Eqs.~(\ref{uni1}) and (\ref{uni2})
and because in our model the matrices $\Gamma_a$ and $\Delta_a$ are diagonal;
in the second line of Eq.~(\ref{sigma-exact}) we also need the 
first Eq.~(\ref{our}) for
$(L_a \hat m R_a^\dagger)_{\mu\tau} = 
(\Gamma_a U_L \hat m U_R^\dagger \Delta_a)_{\mu\tau} =
(\Gamma_a M_D^\dagger \Delta_a)_{\mu\tau} = 0$.
In the definitions of $f_i^a$ and of $g_i^a$ we could,
therefore,
subtract from $\ln{D_i^a}$ its value for $i=1$,
i.e.\
for the lightest neutrino;
this subtraction corresponds to an $i$-independent subtraction
from $f_i^a$ and from $g_i^a$ and,
therefore,
to a null contribution to $\Sigma (p)$.

It is clear from Eqs.~(\ref{G3a}) and (\ref{G4a})
that the relevant $p^2$ in $\Sigma (p)$ is either $m_\mu^2$ or $m_\tau^2$.
When $i = 1,2,3$ one has $m_i^2 \ll p^2 \ll m_a^2$;
when $i = 4,5,6$ one has $p^2 \ll m_a^2 \ll m_i^2$.
It is clear that,
for $i = 1,2,3$,
the logarithm of $D_i^a / D_1^a$
is---neglecting $p^2 \ll m_a^2$---approximately proportional to
$\left( m_i^2 - m_1^2 \right) / m_a^2
\sim m_D^4 / \left( m_R^2 m_F^2 \right)$.
The same happens with the logarithm of $D_i^W / D_1^W$.
Therefore,
the contributions of the light neutrinos
to $\Sigma (p)$ are suppressed by $m_D^4 / \left( m_R^2 m_F^2 \right)$.
On the other hand,
some contributions of the heavy neutrinos remain unsuppressed.
For $i = 4,5,6$ one has
\ba
f_i^a &\approx& \ln{\frac{m_i^2}{m_a^2}}\, ,
\\
g_i^a &\approx& \frac{1}{2} \left( \ln{\frac{m_i^2}{m_a^2}} - 1 \right)\, .
\ea
Matrix elements of $U_L$,
as in $x_i$ and in $( L_a )_{\mu i} ( L_a^\dagger )_{i \tau}$,
suppress some of the contributions of the heavy neutrinos.
Overall one obtains the unsuppressed part 
\ba
\Sigma (p) &\approx&
\frac{1}{16 \pi^2}\, \sum_{i=4}^6 \sum_a
\ln{\frac{m_i^2}{m_a^2}} \left\{
\frac{\, \, \psl \gamma_L}{2}
\left( R_a \right)_{\mu i} \left( R_a^\dagger \right)_{i\tau}
\right. \no & & \left.
- m_i \left[
\left( L_a \right)_{\mu i} \left( R_a^\dagger \right)_{i\tau} \gamma_L
+ \left( R_a \right)_{\mu i} \left( L_a^\dagger \right)_{i\tau} \gamma_R
\right] \right\} .
\label{mvkjd}
\ea
Note that 
the sum over $a$ includes a contribution from the charged Goldstone boson.
\begin{figure}
\begin{center}
\mbox{\epsfig{file=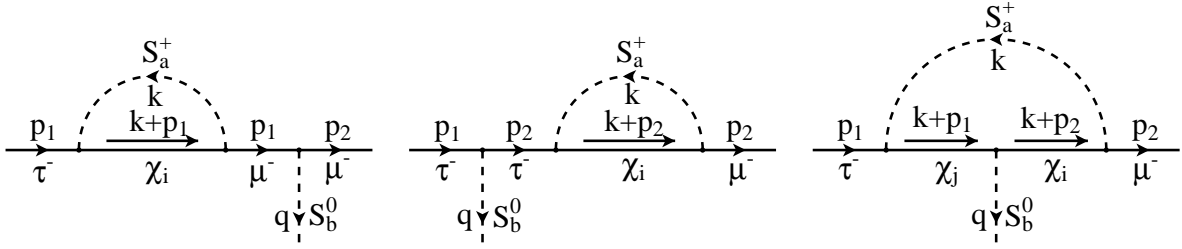,width=0.95\textwidth,height=3.5cm}}
\end{center}
\caption{The three diagrams for $\tau^- \to \mu^- {S_b^0}^*$
which have parts unsuppressed by $m_R^{-1}$.\label{unsuppressed}}
\end{figure}

We conclude that some contributions of the heavy neutrinos
to $M_{b1}$ and to $M_{b2}$ remain unsuppressed when $m_R \to \infty$.
Let us compute those contributions in detail.
Using
\be
m_i \left( U_L \right)_{\ell i}
= \left( M_D^\ast \right)_{\ell \ell} \left( U_R \right)_{\ell i}\, ,
\ee
which is valid in our specific seesaw model for any $\ell $ and $i$,
we rewrite Eq.~(\ref{mvkjd}) as
\ba
\Sigma (p) &\approx&
\frac{1}{16 \pi^2}\, \sum_{i=4}^6 \left[
\left( U_R \right)_{\mu i} \left( U_R^\dagger \right)_{i\tau}
\ln{\frac{m_i^2}{\mu^2}} \right]
\sum_a \left[ \frac{\, \, \psl \gamma_L}{2}
\left( \Delta_a^\ast \right)_{\mu \mu} \left( \Delta_a \right)_{\tau \tau}
\right. \no & & \left.
- \left( \Gamma_a \right)_{\mu \mu}
\left( M_D^\ast \right)_{\mu \mu} \left( \Delta_a \right)_{\tau \tau}
\gamma_L
- \left( \Delta_a^\ast \right)_{\mu \mu} \left( M_D \right)_{\tau \tau}
\left( \Gamma_a^\ast \right)_{\tau \tau}
\gamma_R
\right]
\no &=&
\frac{1}{16 \pi^2}\, \sum_{i=4}^6 \left[
\left( U_R \right)_{\mu i} \left( U_R^\dagger \right)_{i\tau}
\ln{\frac{m_i^2}{\mu^2}} \right]
\sum_k \left[ \frac{\, \, \psl \gamma_L}{2}
\left( \Delta_k^\ast \right)_{\mu \mu} \left( \Delta_k \right)_{\tau \tau}
\right. \no & & \left.
- \left( \Gamma_k \right)_{\mu \mu}
\left( M_D^\ast \right)_{\mu \mu} \left( \Delta_k \right)_{\tau \tau}
\gamma_L
- \left( \Delta_k^\ast \right)_{\mu \mu} \left( M_D \right)_{\tau \tau}
\left( \Gamma_k^\ast \right)_{\tau \tau}
\gamma_R
\right],
\label{approx-sigma}
\ea
where $\mu$ is an arbitrary mass, inserted for dimensional reasons; 
the expression~(\ref{approx-sigma}) is independent of it.
Inserting this result into Eqs.~(\ref{G3a}) and (\ref{G4a}),
we obtain---see Fig.~\ref{unsuppressed}
for the relevant self-energy graphs---
\be
M_{b1} + M_{b2} \approx {\displaystyle \frac{- i}{16 \sqrt{2} \pi^2}\,
\frac{\sum_{i=4}^6 \left[
\left( U_R \right)_{\mu i} \left( U_R^\dagger \right)_{i\tau}
\ln \left( m_i^2/\mu^2 \right) \right]}{m_\tau^2 - m_\mu^2}}\,
\bar u_\mu \left( A_L \gamma_L + A_R \gamma_R \right) u_\tau\, ,
\ee
with
\ba
A_L &=&
\left[
m_\tau \left( \Gamma_b \right)_{\mu\mu}
- m_\mu \left( \Gamma_b \right)_{\tau\tau}
\right]
\left[
\frac{m_\tau}{2}\, \sum_k \left( \Delta_k^\ast \right)_{\mu\mu}
\left( \Delta_k \right)_{\tau\tau}
- \left( M_D \right)_{\tau\tau} \sum_k \left( \Delta_k^\ast \right)_{\mu\mu}
\left( \Gamma_k^\ast \right)_{\tau\tau}
\right]
\no & &
+
\left[
m_\tau \left( \Gamma_b \right)_{\tau\tau}
- m_\mu \left( \Gamma_b \right)_{\mu\mu}
\right]
\left( M_D^\ast \right)_{\mu\mu}
\sum_k \left( \Gamma_k \right)_{\mu\mu} \left( \Delta_k \right)_{\tau\tau}\, ,
\label{AL}
\\
A_R &=&
\left[
m_\tau \left( \Gamma_b^\ast \right)_{\mu\mu}
- m_\mu \left( \Gamma_b^\ast \right)_{\tau\tau}
\right]
\left[
\frac{m_\mu}{2}\, \sum_k \left( \Delta_k^\ast \right)_{\mu\mu}
\left( \Delta_k \right)_{\tau\tau}
- \left( M_D^\ast \right)_{\mu\mu} \sum_k \left( \Gamma_k \right)_{\mu\mu}
\left( \Delta_k \right)_{\tau\tau}
\right]
\no & &
+
\left[
m_\tau \left( \Gamma_b^\ast \right)_{\tau\tau}
- m_\mu \left( \Gamma_b^\ast \right)_{\mu\mu}
\right]
\left( M_D \right)_{\tau\tau}
\sum_k \left( \Delta_k^\ast \right)_{\mu\mu}
\left( \Gamma_k^\ast \right)_{\tau\tau}\, .
\label{AR}
\ea

\subsection{Graphs in which $S^0_b$ couples to charged bosons}

The necessary couplings are found in Appendices \ref{sub:feynman} and
\ref{sub:scalar}. 
Let us define
\be
\label{Diaa'}
D_i^{aa^\prime} = x m_a^2 + y m_{a^\prime}^2
+ \left( 1 - x - y \right) \left( m_i^2 - x m_\tau^2 - y m_\mu^2 \right)
- x y q^2\, .
\ee
Then,
the graph in which $S^0_b$ couples to (incoming) $S_a^-$
and $S_{a^\prime}^+$ yields
\ba
M_{baa^\prime} &=& \frac{-i}{16\pi^2}\, \sum_i
\sum_{a, a^\prime} C_{a a^\prime b} \,
\int_0^1 \! dx\,\int_0^{1-x} \! dy\
\frac{1}{D_i^{aa^\prime}}
\no & &
\times\, \bar u_\mu \left[
\left( R_{a^\prime} \right)_{\mu i} \left( R_a^\dagger \right)_{i\tau}
\left( x m_\tau \gamma_R + y m_\mu \gamma_L \right)
+
\left( L_{a^\prime} \right)_{\mu i} \left( L_a^\dagger \right)_{i\tau}
\left( x m_\tau \gamma_L + y m_\mu \gamma_R \right)
\right. \no & & \left.
- m_i
\left( R_{a^\prime} \right)_{\mu i} \left( L_a^\dagger \right)_{i\tau}
\gamma_R
- m_i
\left( L_{a^\prime} \right)_{\mu i} \left( R_a^\dagger \right)_{i\tau}
\gamma_L
\right] u_\tau\, .
\label{urtsy}
\ea
The cases in which either $S_a^-$,
or $S_{a^\prime}^+$,
or both,
are charged Goldstone bosons,
are implicitly covered through Eqs.~(\ref{RaW}),
(\ref{LaW}),
and~(\ref{C12}), (\ref{C3}).
The graph in which $S^0_b$ couples to $W^-$ and $S_a^+$ gives
\ba
M_{bWa} &=& \frac{- i g^2}{32 \sqrt{2} \pi^2}\,
\sum_i \left( U_L^\dagger \right)_{i\tau}
\sum_a b^\dagger a\,
\int_0^1 \! dx\,\int_0^{1-x} \! dy
\no & &
\times\, \bar u_\mu \left\{
\frac{\left( R_a \right)_{\mu i} m_i}{D_i^{Wa}}
\left[ \left( x + 1 \right) m_\tau \gamma_R
+ \left( y - 2 \right) m_\mu \gamma_L \right]
\right. \no & &
- \frac{\left( L_a \right)_{\mu i}}{D_i^{Wa}}
\left[ \left( 2 x + y \right) m_\mu m_\tau \gamma_R
+ x \left( 1 + x \right) m_\tau^2 \gamma_L
\right. \no & & \left. \left.
+ y \left( y - 2 \right) m_\mu^2 \gamma_L
+ 2 x \left( y - 2 \right) p_1 \cdot p_2 \gamma_L \right]
- 2 \left( L_a \right)_{\mu i} \gamma_L\,
\ln{\frac{D_i^{Wa}}{D_1^{Wa}}} \right\} u_\tau\, .
\label{kdlsp}
\ea
Similarly,
the graph in which $S^0_b$ couples to $S_a^-$ and $W^+$ contributes
\ba
M_{baW} &=& \frac{- i g^2}{32 \sqrt{2} \pi^2}\,
\sum_i \left( U_L \right)_{\mu i}
\sum_a a^\dagger b\,
\int_0^1 \! dx\,\int_0^{1-x} \! dy
\no & &
\times\, \bar u_\mu \left\{
\frac{\left( R_a^\dagger \right)_{i\tau} m_i}{D_i^{aW}} 
\left[ \left( y + 1 \right) m_\mu \gamma_L
+ \left( x - 2 \right) m_\tau \gamma_R \right]
\right. \no & &
- \frac{\left( L_a^\dagger \right)_{i\tau}}{D_i^{aW}} \left[
\left( 2 y + x \right) m_\mu m_\tau \gamma_L
+ x \left( x - 2 \right) m_\tau^2 \gamma_R
\right. \no & & \left. \left.
+ y \left( 1 + y \right) m_\mu^2 \gamma_R
+ 2 y \left( x - 2 \right) p_1 \cdot p_2 \gamma_R
\right]
- 2 \left( L_a^\dagger \right)_{i\tau} \gamma_R\,
\ln{\frac{D_i^{aW}}{D_1^{aW}}} \right\} u_\tau\, .
\label{jxkrs}
\ea
Finally,
there is a graph with the $S^0_b$ attaching to two $W^\pm$ bosons,
yielding
\be
M_{bWW} = \frac{- i g^3 m_W}{16 \pi^2}\, a_W^\dagger b\,
\sum_i x_i\, \int_0^1 \! dx\,\int_0^{1-x} \! dy \ \bar u_\mu
\frac{x m_\tau \gamma_R + y m_\mu \gamma_L}{D_i^{WW}}\, u_\tau\, .
\label{bWW}
\ee
Any part of the integrals which does not depend on $m_i$ ends up,
upon summation over $i$,
giving a zero contribution to $M_b$.
One may therefore subtract from each integral its value when $m_i = m_1$.
In Eqs.~(\ref{kdlsp}) and 
(\ref{jxkrs})
we have already performed that subtraction in the logarithmic terms.
The infinities which occurred together with the logarithms
in those expressions have been dropped using 
$(L_a U_L^\dagger)_{\mu\tau} = (\Gamma_a)_{\mu\tau} = 0$ in Eq.~(\ref{kdlsp})
and the Hermitian-conjugate relation in Eq.~(\ref{jxkrs}).
The amplitudes (\ref{urtsy}) and (\ref{bWW})
have only finite integrals.

Consider for instance $M_{baa^\prime}$ in Eq.~(\ref{urtsy}).
The term with $R_{a^\prime} R_a^\dagger$
is proportional to $(U_R)_{\mu i} (U_R^\dagger)_{i \tau}$.
For $i=1,2,3$ it is suppressed by $m_D^2 / m_R^2$
from the matrix $U_R$;
it is additionally suppressed by $C_{aa'b}$ and the integral,
giving together a factor $m_D / m_F$.
For $i=4,5,6$ the suppression factor is $m_F m_D / m_R^2$.
 
The term with $L_{a^\prime} L_a^\dagger$
has $(U_L)_{\mu i} (U_L^\dagger)_{i \tau}$.
For $i=1,2,3$ we neglect the neutrino masses $m_i$ in the integral
and use Eq.~(\ref{equilibrium}) to obtain the same order of magnitude
as for the term $R_{a^\prime} R_a^\dagger$.
For $i=4,5,6$ we have suppressions $m_D^2 / m_R^2$
from $U_L$ and $m_F m_D / m_R^2$ from the rest of the term.

The terms with $R_{a^\prime} L_a^\dagger$ and $L_{a^\prime} R_a^\dagger$
have suppressions $m_D/m_R$ from the mixing.
For $i=1,2,3$ there is also $m_i$ and the integral,
yielding together suppressions $m_D^2 / \left( m_R m_F \right)$.
For $i=4,5,6$ $m_i \sim m_R$ and the integral is of order $m_R^{-2}$,
and we correspondingly obtain $m_F / m_R$
in addition to the mixing suppression.

In conclusion,
all contributions to $M_{baa^\prime}$ are suppressed by,
at least,
$m_F m_D / m_R^2$.
In the same way,
one finds that $M_{bWa}$ and $M_{baW}$ are suppressed by $m_D^2 / m_R^2$
while $M_{bWW}$ is suppressed by $m_D^3 / \left( m_R^2 m_F \right)$.

\subsection{Graphs in which $S^0_b$ couples to neutrinos}

We remind the reader of the quantity $D_{ij}^W$
defined in Eq.~(\ref{DijW}),
and of the analogous quantity $D_{ij}^a$.
We find,
for the contribution to $M_b$ in which $S^0_b$ couples to neutrinos,
the result
\be
M_{b4} = \frac{-i \sqrt{2}}{16 \pi^2}\, 
\int_0^1 \! dx \int_0^{1-x} \! dy\ \bar u_\mu \sum_{i,j}
\left( \frac{Z_{ij}^W}{D_{ij}^W} + \sum_a \frac{Z_{ij}^a}{D_{ij}^a}
- 2 \sum_a Z_{ij}^{\prime a} \ln{\frac{D_{ij}^a}{D_{11}^a}}
\right) u_\tau\, .
\label{Mb4}
\ee
Here,
\ba
Z_{ij}^W &=&
g^2 \left( U_L \right)_{\mu i} \left( U_L^\dagger \right)_{j \tau} \left[
\left( 1 - x \right) m_\mu m_j \left( F_b^\dagger \right)_{ij} \gamma_L
+ \left( 1 - y \right) m_i m_\tau \left( F_b \right)_{ij} \gamma_R
\right. \no & & \left.
- y m_j m_\tau \left( F_b^\dagger \right)_{ij} \gamma_R
- x m_\mu m_i \left( F_b \right)_{ij} \gamma_L
\right],
\\
Z_{ij}^a &=&
- m_i m_j \left[
\left( R_a \right)_{\mu i}
\left( F_b^\dagger \right)_{ij}
\left( L_a^\dagger \right)_{j \tau} \gamma_R
+
\left( L_a \right)_{\mu i}
\left( F_b \right)_{ij}
\left( R_a^\dagger \right)_{j \tau} \gamma_L
\right]
\no & &
+ m_i m_\tau \left( 1-y \right) \left[
\left( R_a \right)_{\mu i}
\left( F_b^\dagger \right)_{ij}
\left( R_a^\dagger \right)_{j \tau} \gamma_R
+
\left( L_a \right)_{\mu i}
\left( F_b \right)_{ij}
\left( L_a^\dagger \right)_{j \tau} \gamma_L
\right]
\no & &
- m_i m_\mu x \left[
\left( R_a \right)_{\mu i}
\left( F_b^\dagger \right)_{ij}
\left( R_a^\dagger \right)_{j \tau} \gamma_L
+
\left( L_a \right)_{\mu i}
\left( F_b \right)_{ij}
\left( L_a^\dagger \right)_{j \tau} \gamma_R
\right]
\no & &
+ m_\mu m_j \left( 1-x \right) \left[
\left( L_a \right)_{\mu i}
\left( F_b^\dagger \right)_{ij}
\left( L_a^\dagger \right)_{j \tau} \gamma_R
+
\left( R_a \right)_{\mu i}
\left( F_b \right)_{ij}
\left( R_a^\dagger \right)_{j \tau} \gamma_L
\right]
\no & &
- m_\tau m_j y \left[
\left( L_a \right)_{\mu i}
\left( F_b^\dagger \right)_{ij}
\left( L_a^\dagger \right)_{j \tau} \gamma_L
+
\left( R_a \right)_{\mu i}
\left( F_b \right)_{ij}
\left( R_a^\dagger \right)_{j \tau} \gamma_R
\right]
\no & &
- m_\mu m_\tau \left( 1-x-y \right) \left[
\left( L_a \right)_{\mu i}
\left( F_b^\dagger \right)_{ij}
\left( R_a^\dagger \right)_{j \tau} \gamma_R
+
\left( R_a \right)_{\mu i}
\left( F_b \right)_{ij}
\left( L_a^\dagger \right)_{j \tau} \gamma_L
\right]
\no & &
+ \left[ x \left( 1-x \right) m_\mu^2 + y \left( 1-y \right) m_\tau^2
- x y \left( m_\mu^2 + m_\tau^2 - q^2\right) \right]
\no & & \times \left[
\left( L_a \right)_{\mu i}
\left( F_b^\dagger \right)_{ij}
\left( R_a^\dagger \right)_{j \tau} \gamma_L
+
\left( R_a \right)_{\mu i}
\left( F_b \right)_{ij}
\left( L_a^\dagger \right)_{j \tau} \gamma_R
\right],
\ea
and
\be
Z_{ij}^{\prime a} =
\left( L_a \right)_{\mu i}
\left( F_b^\dagger \right)_{ij}
\left( R_a^\dagger \right)_{j \tau} \gamma_L
+
\left( R_a \right)_{\mu i}
\left( F_b \right)_{ij}
\left( L_a^\dagger \right)_{j \tau} \gamma_R\, .
\ee
In Eq.~(\ref{Mb4}) we have already dropped the infinity occurring together
with the logarithm, because, using all the previous
arguments for the cancellation of terms independent of $m_i$, we find
$\sum_{i,j} Z_{ij}^{\prime a} = 0$.

It is tedious but straightforward to check that
all the terms in $Z_{ij}^W$ and in $Z_{ij}^a$
end up suppressed by $1/m_R^2$ or by higher powers of $m_R^{-2}$.
The same does not happen,
however,
with the terms in $Z^{\prime a}_{ij}$
(which include a contribution from the charged Goldstone boson for $a = a_W$).
Let us write the latter in more detail,
using Eqs.~(\ref{La}),
(\ref{Ra}),
(\ref{Ab}),
and the fact that the matrices $\Gamma_a$,
$\Delta_a$,
and $\Delta_b$ are diagonal:
\ba
2 Z^{\prime a}_{ij} &=&
\left( \Gamma_a \right)_{\mu \mu}
\left( \Delta_a \right)_{\tau \tau}
\left( U_L \right)_{\mu i}
\left( U_L^\dagger \Delta_b^\ast U_R
+ U_R^T \Delta_b^\ast U_L^\ast \right)_{ij}
\left( U_R^\dagger \right)_{j \tau} \gamma_L
\no & & +
\left( \Delta_a^\ast \right)_{\mu \mu}
\left( \Gamma_a^\ast \right)_{\tau \tau}
\left( U_R \right)_{\mu i}
\left( U_R^\dagger \Delta_b U_L
+ U_L^T \Delta_b U_R^\ast \right)_{ij}
\left( U_L^\dagger \right)_{j \tau} \gamma_R\, .
\label{z2}
\ea
Due to Eqs.~(\ref{Vsmall}),
terms unsuppressed by $m_D / m_R$ may arise only
from the first term in the right-hand-side of Eq.~(\ref{z2}),
when $i = 1,2,3$ and $j = 4,5,6$,
and from the third term in the right-hand-side of Eq.~(\ref{z2}),
when $i = 4,5,6$ and $j = 1,2,3$.
In the first case the integral of $\ln{D_{ij}^a}$
is practically $i$-independent,
in the second case it is almost $j$-independent.
One obtains the following contribution to $M_{b4}$, with the
corresponding Feynman diagram depicted in Fig.~\ref{unsuppressed}:
\ba
M_{b4}^a &\approx&
\frac{i}{16 \sqrt{2} \pi^2} \sum_{i=4}^6 \left[
\left( U_R \right)_{\mu i} \left( U_R^\dagger \right)_{i \tau}
\ln{\frac{m_i^2}{\mu^2}} \right]
\bar u_\mu \left[
\left( \Delta_b^\ast \right)_{\mu\mu}
\sum_k \left( \Gamma_k \right)_{\mu\mu}
\left( \Delta_k \right)_{\tau\tau}
\gamma_L
\right. \no & & \left. +
\left( \Delta_b \right)_{\tau\tau} \sum_k
\left( \Delta_k^\ast \right)_{\mu\mu}
\left( \Gamma_k^\ast \right)_{\tau\tau}
\gamma_R
\right] u_\tau\, .
\ea
This expression is independent of the arbitrary mass $\mu$.
$M_{b4}$ is not fully suppressed by powers of $m_R$ since $M_{b4}^a$
becomes constant in the limit $m_R \to \infty$.

\subsection{Unsuppressed terms}

We thus conclude that the vertex $\tau^- \to \mu^- {S_b^0}^\ast$
does not vanish in the limit $m_R \to \infty$.
One has
\ba
M_b^\infty & = & \frac{i}{16 \sqrt{2} \pi^2}\,
\sum_{i=4}^6 \left[
\left( U_R \right)_{\mu i} \left( U_R^\dagger \right)_{i \tau}
\ln{\frac{m_i^2}{\mu^2}} \right]
\no & &
\times \bar u_\mu
\left\{
\left[ \left( \Delta_b^\ast \right)_{\mu\mu}
\sum_k \left( \Gamma_k \right)_{\mu\mu}
\left( \Delta_k \right)_{\tau\tau}
+ \frac{A_L}{m_\mu^2 - m_\tau^2} \right]
\gamma_L
\right. \no & & \left.
+
\left[ \left( \Delta_b \right)_{\tau\tau} \sum_k
\left( \Delta_k^\ast \right)_{\mu\mu}
\left( \Gamma_k^\ast \right)_{\tau\tau}
+ \frac{A_R}{m_\mu^2 - m_\tau^2} \right]
\gamma_R
\right\}
u_\tau\, ,
\label{unsup}
\ea
with $A_L$ and $A_R$ given in Eqs.~(\ref{AL}) and (\ref{AR}). Note
that $M_b^\infty$ is independent of the arbitrary mass parameter $\mu$.

It is interesting to observe that $M_b$ {\em is\/} suppressed
when there is only one scalar doublet.
Indeed,
in that case there is only one physical scalar,
the Higgs boson,
which has $b = 1$.
Moreover,
there is only one matrix 
$\Gamma = \left( \sqrt{2} / v^\ast \right) M_\ell$ and, 
since $\left( \sqrt{2} / v \right) M_D = \Delta$,
we find
\begin{equation}
n_H = 1 \; \Rightarrow \left\{
\begin{array}{ccl}
A_L &=&
\frac{\sqrt{2}}{v^\ast}\, \left( m_\tau^2 - m_\mu^2 \right)
\left( M_D^\ast \right)_{\mu\mu}
\left( \Gamma \right)_{\mu\mu} \left( \Delta \right)_{\tau\tau} \\[1mm]
A_R &=&
\frac{\sqrt{2}}{v}\, \left( m_\tau^2 - m_\mu^2 \right)
\left( M_D \right)_{\tau\tau}
\left( \Delta^\ast \right)_{\mu\mu} \left( \Gamma^\ast \right)_{\tau\tau}
\end{array}
\right\} \; \Rightarrow \; M_b^\infty = 0 \,.
\end{equation}
Thus Eq.~(\ref{unsup}) vanishes in this simple case.

\section{Box diagrams} 
\label{sec:box}

\begin{figure}
\begin{center}
\mbox{\epsfig{file=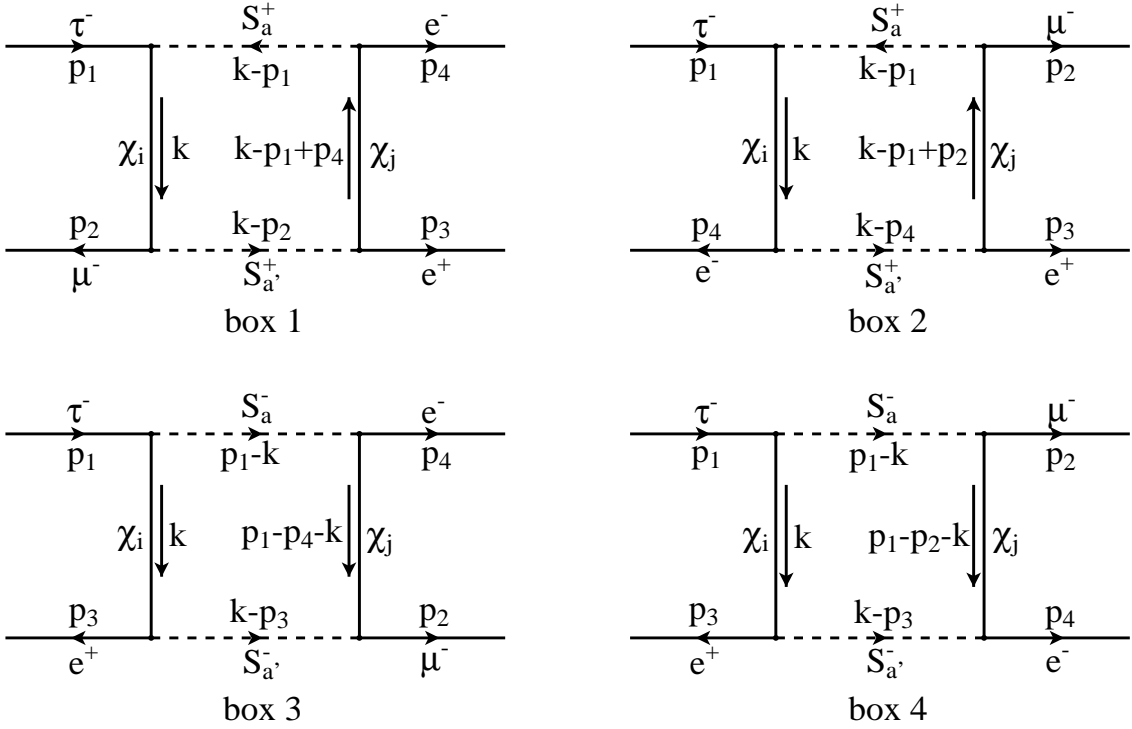,width=0.95\textwidth,height=10cm}}
\end{center}
\caption{The four types of box diagrams for $\tau^- \to \mu^- e^+ e^-$
which occur in our model.\label{box}}
\end{figure}

There are four classes of box diagrams,
as depicted in Fig.~\ref{box}.
In that figure,
either $S_a^\pm$,
or $S_{a^\prime}^\pm$,
or both,
may be substituted by $W^\pm$.

Let us first consider the box diagrams in which
the fermion line starting in the incoming $\tau^-$
ends in the outgoing $\mu^-$.
Those are the diagrams denoted in Fig.~\ref{box} ``box 1''.
The box diagram with two charged scalars $S_a^\pm$ and $S_{a^\prime}^\pm$
gives the following contribution to the decay~(\ref{process}):
\be
\bar M_{aa^\prime} = \frac{i}{16 \pi^2}\,
\int_0^1 \! dx \, \int_0^{1-x} \! dy \, \int_0^{1-x-y}\! dz\,
\sum_{i,j} \sum_{a, a^\prime} \left[
\frac{A_{ij}^{aa^\prime}}{\left( D_{ij}^{aa^\prime} \right)^2}
- \frac{B_{ij}^{aa^\prime}}{2 D_{ij}^{aa^\prime}}
\right].
\label{aaprime}
\ee
The box diagrams with one $S_a^\pm$ and one $W^\pm$ give
\ba
\bar M_{aW} &=& \frac{i g^2}{32 \pi^2}
\sum_{i,j} \left( U_L \right)_{\mu i} \left( U_L^\dagger \right)_{je}\,
\int_0^1 \! dx \int_0^{1-x} \! dy \int_0^{1-x-y}\! dz
\sum_a \left[
\frac{A_{ij}^{aW}}{\left( D_{ij}^{aW} \right)^2}
- \frac{B_{ij}^{aW}}{2 D_{ij}^{aW}}
\right], 
\label{aW} \\
\bar M_{Wa} &=& \frac{i g^2}{32 \pi^2}
\sum_{i,j} \left( U_L^\dagger \right)_{i\tau} \left( U_L \right)_{ej}\,
\int_0^1 \! dx \int_0^{1-x} \! dy \int_0^{1-x-y}\! dz
\sum_a \left[
\frac{A_{ij}^{Wa}}{\left( D_{ij}^{Wa} \right)^2}
- \frac{B_{ij}^{Wa}}{2 D_{ij}^{Wa}}
\right].
\label{Wa}
\ea
\noindent
Finally,
the box diagram with two $W^\pm$ yields
\ba
\bar M_{WW} &=& \frac{i g^4}{64 \pi^2}
\sum_{i,j} \left( U_L \right)_{\mu i} \left( U_L^\dagger \right)_{i\tau}
\left( U_L \right)_{ej} \left( U_L^\dagger \right)_{je}
\no & & \times
\int_0^1 \! dx \, \int_0^{1-x} \! dy \, \int_0^{1-x-y}\! dz\, \left[
\frac{A_{ij}^{WW}}{\left( D_{ij}^{WW} \right)^2}
- \frac{B_{ij}^{WW}}{2 D_{ij}^{WW}}
\right].
\label{WW}
\ea
In Eqs.~(\ref{aaprime})--(\ref{WW}),
\ba
D_{ij}^{aa^\prime} &=& x m_a^2 + y m_{a^\prime}^2 + z m_j^2
+ \left( 1 - x - y - z \right)
\left( m_i^2 - x p_1^2 - y p_2^2 - z r^2 \right)
\no & &
- x y q^2 - x z p_4^2 - y z p_3^2\, ;
\label{Dijaa}
\ea
$D_{ij}^{aW}$,
$D_{ij}^{Wa^\prime}$,
and $D_{ij}^{WW}$ are obtained from $D_{ij}^{aa^\prime}$
by substituting either $m_{a^\prime}$,
or $m_a$,
or both,
by $m_W$.
In order to write down the expressions
for $A_{ij}^{aa^\prime}, \ldots , B_{ij}^{WW}$ one must define
\ba
P^\alpha &=& p_1^\alpha \left( x + z \right)
+ p_2^\alpha y - p_4^\alpha z\, ,
\label{Palfa} \\
P^{\prime\alpha} &=& p_1^\alpha \left( x + z - 1 \right)
+ p_2^\alpha y + p_4^\alpha \left( 1 - z \right).
\label{Pprimealfa}
\ea
One then has
\ba
A_{ij}^{aa^\prime} &=&
\bar u_\mu \left[
\left( R_{a^\prime} \right)_{\mu i} \left( R_a^\dagger \right)_{i \tau}
\!\! \not \! P \gamma_L
+
\left( L_{a^\prime} \right)_{\mu i} \left( L_a^\dagger \right)_{i \tau}
\!\! \not \! P \gamma_R
-
m_i \left( R_{a^\prime} \right)_{\mu i} \left( L_a^\dagger \right)_{i \tau}
\gamma_R
\right. \no & & \left.
-
m_i \left( L_{a^\prime} \right)_{\mu i} \left( R_a^\dagger \right)_{i \tau}
\gamma_L
\right] u_\tau\,
\bar u_e \left[
\left( R_a \right)_{ej} \left( R_{a^\prime}^\dagger \right)_{je}
\!\! \not \! P^\prime \gamma_L
+
\left( L_a \right)_{ej} \left( L_{a^\prime}^\dagger \right)_{je}
\!\! \not \! P^\prime \gamma_R
\right. \no & & \left.
-
m_j \left( R_a \right)_{ej} \left( L_{a^\prime}^\dagger \right)_{je}
\gamma_R
-
m_j \left( L_a \right)_{ej} \left( R_{a^\prime}^\dagger \right)_{je}
\gamma_L
\right] v_e\, ,
\label{Aaaprime}
\\
A_{ij}^{aW} &=&
- \bar u_\mu \gamma^\rho \gamma_L \left[
m_i \left( R_a^\dagger \right)_{i\tau}
- \!\! \not \! P \! \left( L_a^\dagger \right)_{i\tau}
\right] u_\tau\, \bar u_e \left[
\left( R_a \right)_{ej} m_j
- \left( L_a \right)_{ej} \!\! \not \! P^\prime
\right] \gamma_\rho \gamma_L v_e\, ,
\label{AaW} \\
A_{ij}^{Wa} &=&
- \bar u_\mu \left[ \left( R_a \right)_{\mu i} m_i
- \left( L_a \right)_{\mu i} \!\! \not \! P
\right] \gamma^\rho \gamma_L u_\tau\,
\bar u_e \gamma_\rho \gamma_L \left[
m_j \left( R_a^\dagger \right)_{je}
- \!\! \not \! P^\prime \left( L_a^\dagger \right)_{je}
\right] v_e\, , \hspace*{5mm}
\label{AWa}
\\
A_{ij}^{WW} &=&
\bar u_\mu \gamma^\rho \!\! \not \! P
\gamma^\sigma \gamma_L u_\tau\,
\bar u_e \gamma_\sigma \!\! \not \! P^\prime
\gamma_\rho \gamma_L v_e\, .
\label{AWW}
\ea
One obtains $B_{ij}^{aa^\prime}$ by discarding from $A_{ij}^{aa^\prime}$
all the terms which contain neither $\not \! P$ nor $\not \! P^\prime$,
and then substituting,
in the remaining terms,
$\not \! P$ by $\gamma^\theta$
and $\not \! P^\prime$ by $\gamma_\theta$.
Applying the same algorithm,
one obtains $B_{ij}^{aW}$ from $A_{ij}^{aW}$,
$B_{ij}^{Wa}$ from $A_{ij}^{Wa}$,
and $B_{ij}^{WW}$ from $A_{ij}^{WW}$.

Next we consider the box diagrams in which
the fermion line starting in the incoming $\tau^-$
ends in the outgoing $e^-$.
Those are the diagrams called ``box 2'' in Fig.~\ref{box}.
The corresponding amplitudes may be obtained
from Eqs.~(\ref{aaprime})--(\ref{AWW}) by using,
instead of Eqs.~(\ref{Dijaa})--(\ref{Pprimealfa}),
\ba
D_{ij}^{aa^\prime} &=& x m_a^2 + y m_{a^\prime}^2 + z m_j^2
+ \left( 1 - x - y - z \right)
\left( m_i^2 - x p_1^2 - y p_4^2 - z q^2 \right)
\no & &
- x y r^2 - x z p_2^2 - y z p_3^2\, ,
\\
P^\alpha &=& p_1^\alpha \left( x + z \right)
- p_2^\alpha z + p_4^\alpha y\, ,
\\
P^{\prime\alpha} &=& p_1^\alpha \left( x + z - 1 \right)
+ p_2^\alpha \left( 1 - z \right) + p_4^\alpha y\, ,
\ea
and by making,
in Eqs.~(\ref{Aaaprime})--(\ref{AWW}),
the substitutions $\bar u_\mu \leftrightarrow \bar u_e$,
$\left( L_a, R_a \right)_{ej}
\rightarrow \left( L_a, R_a \right)_{\mu j}$,
and
$\left( L_{a,a^\prime}, R_{a,a^\prime} \right)_{\mu i}
\rightarrow \left( L_{a,a^\prime}, R_{a,a^\prime} \right)_{ei}$.
Finally,
one must change the overall sign of the amplitudes,
i.e.\ insert a minus sign in front
of Eqs.~(\ref{aaprime})--(\ref{WW}),
due to the interchange of two fermions in the final state.
In Eqs.~(\ref{aW})--(\ref{WW}) one must also interchange
$\left( U_L \right)_{\mu i} \rightarrow \left( U_L \right)_{ei}$
and $\left( U_L \right)_{ej} \rightarrow \left( U_L \right)_{\mu j}$.

Another type of diagrams are 
the box diagrams denoted ``box 3'' in Fig.~\ref{box}.
Those diagrams arise due to the Majorana character of the neutrinos,
and they must be computed using
specific Feynman rules for Majorana fields---see, for instance,
Ref.~\cite{Feynman-Majorana}.
One obtains,
analogously to Eqs.~(\ref{aaprime})--(\ref{WW}),
the following contributions to the decay~(\ref{process}):
\ba
\tilde M_{aa^\prime} &=& \frac{i}{16 \pi^2}\,
\int_0^1 \! dx \, \int_0^{1-x} \! dy \, \int_0^{1-x-y}\! dz\,
\sum_{i,j} \sum_{a, a^\prime} \left[
\frac{\tilde A_{ij}^{aa^\prime}}{\left( \tilde D_{ij}^{aa^\prime} \right)^2}
- \frac{\tilde B_{ij}^{aa^\prime}}{2 \tilde D_{ij}^{aa^\prime}}
\right],
\label{jxlep} \\
\tilde M_{aW} &=& \frac{i g^2}{32 \pi^2}
\sum_{i,j} \left( U_L^\dagger \right)_{ie} \left( U_L \right)_{\mu j}\,
\int_0^1 \! dx \int_0^{1-x} \! dy \int_0^{1-x-y}\! dz
\sum_a \left[
\frac{\tilde A_{ij}^{aW}}{\left( \tilde D_{ij}^{aW} \right)^2}
- \frac{\tilde B_{ij}^{aW}}{2 \tilde D_{ij}^{aW}}
\right],
\no & &
\\
\tilde M_{Wa} &=& \frac{i g^2}{32 \pi^2}
\sum_{i,j} \left( U_L^\dagger \right)_{i\tau} \left( U_L \right)_{ej}\,
\int_0^1 \! dx \int_0^{1-x} \! dy \int_0^{1-x-y}\! dz
\sum_a \left[
\frac{\tilde A_{ij}^{Wa}}{\left( \tilde D_{ij}^{Wa} \right)^2}
- \frac{\tilde B_{ij}^{Wa}}{2 \tilde D_{ij}^{Wa}}
\right],
\no & &
\\
\tilde M_{WW} &=& \frac{i g^4}{64 \pi^2}
\sum_i m_i \left( U_L^\dagger \right)_{ie}
\left( U_L^\dagger \right)_{i\tau}
\sum_j m_j \left( U_L \right)_{ej} \left( U_L \right)_{\mu j}
\no & & \times
\int_0^1 \! dx \, \int_0^{1-x} \! dy \, \int_0^{1-x-y}\! dz\, \left[
\frac{\tilde A_{ij}^{WW}}{\left( \tilde D_{ij}^{WW} \right)^2}
- \frac{\tilde B_{ij}^{WW}}{2 \tilde D_{ij}^{WW}}
\right].
\label{ucfak}
\ea
Here,
\ba
\tilde D_{ij}^{aa^\prime} &=& x m_a^2 + y m_{a^\prime}^2 + z m_j^2
+ \left( 1 - x - y - z \right) m_i^2
\no & &
- x \left( 1 - x - z \right) p_1^2
- y \left( x + z \right) p_2^2
- y \left( 1 - y - z \right) p_3^2
- x \left( y + z \right) p_4^2
\no & &
+ x y q^2
+ \left[ x y - z \left( 1 - x - y - z \right) \right] r^2\, .
\label{kfjwl}
\ea
Notice that $\tilde M_{WW}$ is  proportional to $m_i$ and to $m_j$,
indicating that it vanishes in both limits $m_i \to 0$ and $m_j \to 0$.
This is an instance
of Kayser's ``practical Majorana--Dirac confusion theorem'' \cite{kayser}.
Defining
\ba
\tilde P^\alpha &=& p_1^\alpha \left( x + z \right)
+ p_3^\alpha y - p_4^\alpha z\, ,
\\
\tilde P^{\prime\alpha} &=& p_1^\alpha \left( x + z - 1 \right)
+ p_3^\alpha y + p_4^\alpha \left( 1 - z \right)\, ,
\label{ocurm}
\ea
one may write
\ba
\tilde A_{ij}^{aa^\prime} &=&
v_e^T C^{-1} \left[
\left( R_{a^\prime}^\dagger \right)_{ie} \left( L_a^\dagger \right)_{i\tau}
\!\! \not \! \tilde P \gamma_R
+ \left( L_{a^\prime}^\dagger \right)_{ie} \left( R_a^\dagger \right)_{i\tau}
\!\! \not \! \tilde P \gamma_L
- m_i \left( R_{a^\prime}^\dagger \right)_{ie}
\left( R_a^\dagger \right)_{i\tau} \gamma_L
\right. \no & & \left.
- m_i \left( L_{a^\prime}^\dagger \right)_{ie}
\left( L_a^\dagger \right)_{i\tau} \gamma_R
\right] u_\tau\,
\bar u_\mu \left[
\left( R_{a^\prime} \right)_{\mu j}
\left( L_a \right)_{ej} \!\! \not \! \tilde P^\prime \gamma_L
+ \left( L_{a^\prime} \right)_{\mu j}
\left( R_a \right)_{ej} \!\! \not \! \tilde P^\prime \gamma_R
\right. \no & & \left.
+ m_j \left( R_{a^\prime} \right)_{\mu j}
\left( R_a \right)_{ej} \gamma_R
+ m_j \left( L_{a^\prime} \right)_{\mu j}
\left( L_a \right)_{ej} \gamma_L
\right] C \bar u_e^T\, ,
\label{uttds}
\\
\tilde A_{ij}^{aW} &=&
v_e^T C^{-1} \gamma^\rho \gamma_R
\left[ \left( R_a^\dagger \right)_{i\tau} \!\! \not \! \tilde P
- m_i \left( L_a^\dagger \right)_{i\tau} \right] u_\tau\,
\bar u_\mu \gamma_\rho \gamma_L \left[
\left( R_a \right)_{ej} \!\! \not \! \tilde P^\prime
+ \left( L_a \right)_{ej} m_j
\right] C \bar u_e^T\, ,
\no & &
\\
\tilde A_{ij}^{Wa} &=&
v_e^T C^{-1}
\left[ \left( R^\dagger_a \right)_{ie} \!\! \not \! \tilde P
- m_i \left( L^\dagger_a \right)_{ie} \right]
\gamma^\rho \gamma_L u_\tau\,
\bar u_\mu \left[
\left( R_a \right)_{\mu j} \!\! \not \! \tilde P^\prime
+ \left( L_a \right)_{\mu j} m_j
\right] \gamma_\rho \gamma_R C \bar u_e^T\, ,
\no & &
\\
\tilde A_{ij}^{WW} &=&
- v_e^T C^{-1} \gamma^\rho \gamma^\sigma \gamma_L u_\tau\,
\bar u_\mu \gamma_\rho \gamma_\sigma \gamma_R C \bar u_e^T\, .
\label{jdlks}
\ea
One obtains $\tilde B_{ij}^{aa^\prime}$
from $\tilde A_{ij}^{aa^\prime}$---and similarly 
$\tilde B_{ij}^{aW}$ from $\tilde A_{ij}^{aW}$,
$\tilde B_{ij}^{Wa}$ from $\tilde A_{ij}^{Wa}$,
and $\tilde B_{ij}^{WW}$ from $\tilde A_{ij}^{WW}$---by deleting
all the terms which contain neither $\not \! \tilde P$
nor $\not \! \tilde P^\prime$,
and then substituting,
in the remaining terms,
$\not \! \tilde P$ by $\gamma^\theta$
and $\not \! \tilde P^\prime$ by $\gamma_\theta$.

Finally,
there are the box diagrams of the type denoted in Fig.~\ref{box} ``box 4''.
The corresponding amplitudes may be obtained
from Eqs.~(\ref{jxlep})--(\ref{jdlks}) in the following way.
Firstly,
one must change the overall sign
of Eqs.~(\ref{jxlep})--(\ref{ucfak}) and perform in them the interchange
$\left( U_L \right)_{ej} \leftrightarrow \left( U_L \right)_{\mu j}$.
Secondly,
in Eqs.~(\ref{kfjwl})--(\ref{ocurm})
one must interchange $p_2^\alpha \leftrightarrow p_4^\alpha$,
$p_2^2 \leftrightarrow p_4^2$,
and $q^2 \leftrightarrow r^2$.
Lastly,
in Eqs.~(\ref{uttds})--(\ref{jdlks}) one must interchange
$\bar u_\mu \leftrightarrow \bar u_e$ and
$\left( L_{a,a^\prime}, R_{a,a^\prime} \right)_{ej} \leftrightarrow
\left( L_{a,a^\prime}, R_{a,a^\prime} \right)_{\mu j}$.

One may analyze the $m_R$ dependence of the box amplitudes given above
by using the skills developed in the previous sections.
It is easily concluded that all those amplitudes
are suppressed by at least one factor $m_R^{-2}$.
In order to reach this conclusion,
it is necessary to use Eq.~(\ref{equilibrium}),
when the box diagrams are either of type ``box 1'' or ``box 2''
and both $\chi_i$ and $\chi_j$ are light neutrinos.

\section{The asymptotic limit $\lowercase{m}_R \to \infty$}
\label{sec:asymptotic}

In our model there is a scale $m_R$
which is much higher than the other two scales,
$m_D$ and $m_F$ (see Section \ref{sec:orders}).
In this section we want to study the asymptotic limit $m_R \to \infty$
of our model.
The simplest way---the one which we have in mind
in the following---of increasing the scale $m_R$
is by multiplying the mass matrix $M_R$ in Eq.~(\ref{MR})
by a (dimensionless) factor which becomes much larger than one.

When the decoupling theorem \cite{appelquist,collins}
applies straightforwardly,
one may simply delete the heavy fields from the Lagrangian
in order to obtain the low-energy theory.
In our case this does not work because,
if we remove the fields $\nu_R$ from the Lagrangian,
we also delete any trace of the flavor-changing neutral interactions.
This is at odds with the explicit one-loop calculation
of the vertex $\ell \to \ell^\prime {S_b^0}^\ast$,
for $\ell \neq \ell^\prime$,
since that vertex does not vanish in the limit $m_R \to \infty$
(see Fig.~\ref{unsuppressed} for the Feynman diagrams
with non-vanishing contributions). 
On the other hand,
according to Ref.~\cite{collins},
the limit $m_R \to \infty$ {\em must\/} yield a sensible theory.
Evidently,
the only theory which can emerge from our model in that limit
is the multi-Higgs-doublet SM,
containing flavor-changing neutral Higgs interactions
suppressed by loop factors $16 \pi^2$ and by small couplings.
Thus,
we expect a Yukawa Lagrangian of the form
\begin{equation}\label{YY}
\tilde{\mathcal{L}}_\mathrm{Y} = - \sum_{j=1}^{n_H}
\left( 
\begin{array}{cc} \varphi_j^-\,, & {\varphi_j^0}^\ast \end{array}
\right)
\bar{\ell}_R\, \tilde{\Gamma}_j \left(
\begin{array}{c} \tilde{\nu}_L \\ \ell_L \end{array}
\right) + \mathrm{H.c.}
\end{equation}
In order to demonstrate that this Lagrangian indeed emerges,
we consider the following couplings:
\begin{enumerate}
\item
$\ell_L \to \ell^\prime_R {S^0}^\ast$,
\item
$\nu_L \to \ell_R {S^+}^\ast$,
\item
$\nu_L \to \ell_L {W^+}^\ast$.
\end{enumerate}
For the sake of brevity,
let us denote the corresponding tree-level vertices by $V_0$,
$V_+$,
and $V_W$, 
respectively.
The one-loop contributions to these vertices fall into three categories:
\begin{enumerate}
\renewcommand{\theenumi}{\Roman{enumi}}
\item
The contribution has a non-vanishing limit $m_R \to \infty$.
\item 
In the limit $m_R \to \infty$ the contribution vanishes.
\item 
The contribution is independent of $m_R$ and is present also
when the fields $\nu_R$ are removed from the Lagrangian.
\end{enumerate}
Our strategy is the following.
We identify all contributions of Category I,
since only they are relevant for obtaining the limit $m_R \to \infty$
of our model.
We then show that there are three contributions of Category I to $V_W$,
but they cancel out except for a part which may be viewed
as a unitary transformation on the vector of flavor fields $\nu_L$.
The transformed field vector is denoted by $\tilde\nu_L$.
After these steps,
we see that the contributions of Category I
to the vertices $V_0$ and $V_+$ are identical,
provided that the neutrino field $\tilde\nu_L$ is used in $V_+$.
This demonstrates that in the limit $m_R \to \infty$
the Yukawa Lagrangian of Eq.~(\ref{YY}) emerges. 
After the one-loop corrections,
as discussed above,
we remove the fields $\nu_R$ from the Lagrangian.
Then all the traces left by those heavy fields
are contained in the off-diagonal elements of $\tilde \Gamma_j$,
which have arisen from the one-loop corrections of Category I.

We now pursue the strategy outlined in the previous paragraph.
First consider the vertex
and self-energy corrections for the above couplings: 
\begin{eqnarray}
V_0: &&
\Gamma_k + \Delta \Gamma_k \left( V_0,S^\pm \right) + 
\frac{1}{2} \left[ z_R^\ell \left( S^\pm \right) \right]^\dagger \Gamma_k +
\frac{1}{2}\, \Gamma_k z_L^\ell \left( S^\pm \right), \label{V0} \\
V_+: &&
\Gamma_k + \Delta \Gamma_k \left( V_+,S^0 \right) + 
\frac{1}{2} \left[ z_R^\ell \left( S^\pm \right) \right]^\dagger \Gamma_k +
\frac{1}{2}\, \Gamma_k z_L^\nu \left( S^0 \right) \,, \label{V+} \\
V_W: &&
\frac{g}{\sqrt{2}} \left\{ 1 + \Gamma \left( V_W,S \right) +
\frac{1}{2} \left[ z_L^\ell \left( S^\pm \right) \right]^\dagger +
\frac{1}{2}\, z_L^\nu \left( S^0 \right) \right\}. \label{VW}
\end{eqnarray}
For the definition of the wave-function renormalization matrices $z_L^\ell$,
$z_R^\ell$,
and $z_L^\nu$ see Appendix~\ref{app:renormalization}.
It is proven by a tedious checking of all one-loop
graphs that the corrections to $V_0$,
$V_+$,
and $V_W$ not included in Eqs.~(\ref{V0}),
(\ref{V+}),
and (\ref{VW}),
respectively,
are either of Category II or Category III.
In the expressions (\ref{V+}) and (\ref{VW}),
only one-loop corrections
in which the external neutrino legs correspond to light neutrinos
are interesting.
Those one-loop corrections are obtained
by performing the calculations with the fields $\chi_i$,
but the result has to be multiplied appropriately by $U_L$;
in order to obtain $\Delta \Gamma_k (V_+,S^\pm )$
a multiplication from the right by $U_L^\dagger$ is necessary;
the $6 \times 6$ matrix part of the neutrino self-energy
associated with the Dirac structure $\ppsl \gamma_L$
has to be multiplied by $U_L$ from the left and $U_L^\dagger$ from the right
in order to arrive at the $3 \times 3$ matrix we need; and so on.
With this procedure we have in mind that,
in the limit $m_R \to \infty$,
the light neutrinos become massless,
and we are allowed to work with the fields $\nu_L$,
which are members of the left-handed lepton doublets
in the Lagrangian before spontaneous symmetry breaking.
In the following,
our results for the quantities appearing in Eqs.~(\ref{V0}),
(\ref{V+}),
and (\ref{VW}) will only be given for $m_R \to \infty$,
since this is the limit we aim at.
In Eq.~(\ref{V0}) the vertex correction $\Delta \Gamma_k(V_0,S^\pm)$
is obtained by exchanging all charged scalars,
including the charged Goldstone boson $G^\pm$.
In Eq.~(\ref{V+}) the vertex correction $\Delta \Gamma_k(V_+,S^0)$
stems from exchanging all neutral scalars,
including the neutral Goldstone boson $G^0$.
The vertex correction $\Gamma(V_W,S)$
originates in the coupling of the W boson
to a charged and a neutral scalar;
thus,
the loop has $S^0_b$,
$S^\pm_a$,
and $\chi_i$.
Concerning the wave-function renormalization matrices $z_L^\ell(S^\pm)$,
$z_R^\ell(S^\pm)$,
and $z_L^\nu(S^0)$,
in parentheses we have indicated the scalar exchange
they come from in the self-energies.

In Eqs.~(\ref{V0}),
(\ref{V+}),
and (\ref{VW}),
all the one-loop contributions which occur
contain two $\Delta$ and one $\Gamma$ Yukawa-coupling matrices.
The vertices $V_0$ and $V_+$ also receive contributions
from the wave-function renormalization matrix of the Higgs doublets;
these contributions fall into all three categories. One can show that
those of Category I induce the same corrections to $\Gamma_k$ at the
vertices $V_0$ and $V_+$.

Before we list the one-loop results
for the quantities appearing in Eqs.~(\ref{V0}),
(\ref{V+}),
and (\ref{VW}),
we want to introduce some useful notation.
First we note that
\begin{equation}
\lim_{m_R \to \infty} U_R = (0, W) \,, \quad
\lim_{m_R \to \infty} m_{1,2,3} = 0 \,,
\end{equation}
where $W$ is a $3 \times 3$ unitary matrix. 
We define a useful $3 \times 3$ Hermitian matrix by
\begin{equation}\label{A}
A \equiv \frac{1}{16\pi^2} \sum_{j=1}^{n_H} \Delta_j^\dagger  
\left[ -\frac{1}{2}\, \mathcal{K} - \frac{3}{4} + \frac{1}{2} \,
W \ln {\tilde{m}}^2\, W^\dagger \right] \Delta_j \,,
\end{equation}
where the divergent constant $\mathcal{K}$ has been defined in Eq.~(\ref{K})
and
\begin{equation}
\tilde m = \mathrm{diag} \left( m_4, m_5, m_6 \right) .
\end{equation}
Then,
it is easy to write down the self-energy
\begin{equation}\label{sigmanu}
\Sigma_\nu^\infty \left( S^0;p \right) = A \ppsl \gamma_L
\end{equation}
for the neutrinos,
and
\begin{equation}\label{sigmaell}
\Sigma_\ell^\infty \left( S^\pm;p \right) = 
A \ppsl \gamma_L + B \gamma_L + B^\dagger \gamma_R 
\end{equation}
for the charged leptons,
with $B$ given by
\begin{equation}\label{B}
B = \frac{1}{16\pi^2} \sum_{j=1}^{n_H}
\Gamma_j \left[ M_D^\dagger \left( \mathcal{K} + 1 \right)
- W \tilde m \ln{\tilde m^2} W^\dagger \right] \Delta_j\, .
\end{equation}
The superscript $\infty$ reminds us that we have taken
the limit $m_R \to \infty$.
In the same limit,
the vertex corrections are given by
\begin{equation}\label{DG}
\Delta \Gamma_k \left( V_0,S^\pm \right)
= \Delta \Gamma_k \left( V_+,S^0 \right)
= \frac{1}{16\pi^2} \sum_{j=1}^{n_H} \Gamma_j \Delta_k^\dagger 
\left( \mathcal{K} + 1 - W \ln{\tilde m^2} W^\dagger \right) \Delta_j
\end{equation}
and
\begin{equation}\label{VWA}
\Gamma \left( V_W,S \right) = - A\, .
\end{equation}
The off-diagonal elements of $\Delta \Gamma_k(V_0,S^\pm)$
are given by the vertex correction computed in Section~\ref{sec:S},
in the limit $m_R \to \infty$;
the same holds for $\Sigma_\ell^\infty(S^\pm;p)$.
In order to calculate the self-energy (\ref{sigmanu})
one has to take into account the Majorana nature of the fields $\chi_i$:
the neutrino self-energy derives from a propagator
at second order in perturbation theory,
and there are $2^3 = 8$ possibilities
to attach the external legs to neutrino fields
in the two $S^0$ Yukawa Lagrangians;
therefore there is a combinatorial factor $2^3 / 2! = 4$.
Furthermore,
the terms $B_{L,R}$ drop out of the neutrino self-energy because of
\cite{grimus89}
\begin{equation}
\sum_b b_k b_{k^\prime} = 0 \,.
\end{equation}
We have also used the relations (see Appendix A.1)
\begin{equation}
\sum_b b_k^\ast b_{k^\prime}
= 2 \sum_a a_k^\ast a_{k^\prime} = 2 \delta_{k k^\prime}\, ,
\end{equation}
which have enabled us to sum,
in Eqs.~(\ref{A}),
(\ref{B}) and (\ref{DG}) over the index $j$.

Since in the limit $m_R \to \infty$ the neutrinos are massless,
the procedure for the renormalization of the self-energy
laid out in Appendix B.1 is not applicable.
However,
it is reasonable,
in view of Eq.~(\ref{sigmanu}),
to define
\begin{equation}\label{znu}
z_R^\nu \left( S^0 \right) = 0\, , \quad 
z_L^\nu \left( S^0 \right)
= \left[ z_L^\nu \left( S^0 \right) \right]^\dagger = A \,.
\end{equation}
The determination of $z_L^\ell$ and of $z_R^\ell$
follows the on-shell prescription in Appendix B.1.

In the expression for the one-loop correction of $V_W$,
Eq.~(\ref{VW}),
the only matrix which is possibly non-Hermitian is $z_L^\ell(S^ \pm)$.
In any case,
we may decompose that matrix into a Hermitian and an anti-Hermitian part:
\begin{equation}
z_L^\ell \left( S^\pm \right) = z_+^\ell + z_-^\ell\, ,
\quad \mathrm{with} \quad
\left( z_\pm^\ell \right)^\dagger = \pm z_\pm^\ell \,.
\end{equation}
We notice that in the $\Sigma_\ell^\infty(S^\pm;p)$
of Eq.~(\ref{sigmaell}) there is no term $A_R^\ell \ppsl$.
Using the equations in Appendix~B.1 for calculating
the fermion wave-function renormalization matrices,
this fact simplifies considerably the expression for $z^\ell_{L,R}$.
Since $A$ is Hermitian,
we obtain
\begin{equation}
\mathrm{Re} \left[ z_L^\ell \left( S^\pm \right)
- z_R^\ell \left( S^\pm \right) \right]_{\alpha\alpha} =
A_{\alpha\alpha} \,, \quad
\mathrm{Im} \left[ z_L^\ell \left( S^\pm \right)
- z_R^\ell \left( S^\pm \right) \right]_{\alpha\alpha} =
- \frac{2}{m_\alpha} \mathrm{Im}\, B_{\alpha\alpha}\, ,
\end{equation}
and
\begin{equation}
\mathrm{Re} \left[ z_L^\ell \left( S^\pm \right)
+ z_R^\ell \left( S^\pm \right) \right]_{\alpha\alpha} =
A_{\alpha\alpha} \,.
\end{equation}
Using the possibility to choose the convention
\begin{equation}\label{convention}
\mathrm{Im} \left[ z_L^\ell \left( S^\pm \right) \right]_{\alpha\alpha} = 0 \,,
\end{equation}
as explained at the end of Appendix \ref{sub:onshell},
we obtain 
$[ z_L^\ell(S^\pm) ]_{\alpha\alpha} = ( z_+^\ell )_{\alpha\alpha} = 
A_{\alpha\alpha}$
and $( z_-^\ell )_{\alpha\alpha} = 0$. 
For $\alpha \neq \beta$ we apply again
the procedure of Appendix B.1 and obtain
$( z_+^\ell )_{\beta\alpha} = A_{\beta\alpha}$.
In summary,
with the convention of Eq.~(\ref{convention}) we find
\begin{equation}\label{zLA}
z_+^\ell = A 
\end{equation}
and,
with Eqs.~(\ref{VWA}) and (\ref{znu}),
\begin{equation}\label{cancel}
\Gamma \left( V_W,S \right)
+ \frac{1}{2} \left( z_+^\ell + z_L^\nu \right) = 0 \,.
\end{equation}
Thus, Eq.~(\ref{VW}) is given by
\begin{equation}\label{VWcancel}
\frac{g}{\sqrt{2}} \left( 1 - \frac{1}{2} z_-^\ell \right).
\end{equation}
Since the light-neutrino masses vanish when $m_R \to \infty$,
we are allowed to rotate $\nu_L$ unitarily.
An infinitesimal unitary rotation is given by
\begin{equation}
\nu_L = \left( 1+\Omega \right) {\tilde \nu}_L\, , 
\quad \mathrm{with} \quad \Omega^\dagger = -\Omega \,.
\end{equation}
With the choice \cite{kniehl}
\begin{equation}
\Omega = \frac{1}{2}\, z_-^\ell \,,
\end{equation}
the one-loop-corrected W vertex of Eq.~(\ref{VW}) 
reduces to the trivial form $g/\sqrt{2}$,
if conceived as pertaining to ${\tilde \nu}_L$.

In terms of the new field $\tilde \nu_L$,
the vertex $V_+$ is given by
\begin{equation}\label{V++}
\Gamma_k + \Delta \Gamma_k \left( V_+,S^0 \right) + 
\frac{1}{2} \left[ z_R^\ell \left( S^\pm \right) \right]^\dagger \Gamma_k +
\frac{1}{2}\, \Gamma_k \left[ z_L^\nu \left( S^0 \right)
+ z_-^\ell \right].
\end{equation}
Since
\begin{equation}
z_L^\nu \left( S^0 \right) + z_-^\ell
= A + z_-^\ell = z_L^\ell \left( S^\pm \right),
\end{equation}
and because of Eq.~(\ref{DG}),
the expression (\ref{V++}) for the charged-scalar vertex 
is identical with the expression (\ref{V0}) for the neutral-scalar vertex.
Therefore,
in the $\tilde \nu_L$ basis for the neutrino fields,
the one-loop corrections---associated with $\nu_R$---to
the couplings of types $\bar\ell_R \ell_L S^0$
and $\bar\ell_R \tilde\nu_L S^-$ are identical.
This is crucial for writing the theory after decoupling
of the heavy fields $\nu_R$
as a multi-Higgs-doublet Standard Model
with the Yukawa Lagrangian of Eq.~(\ref{YY}).
The coupling matrices $\tilde \Gamma_k$ are then given by Eq.~(\ref{V0}).
The infinities introduced by the one-loop contributions of Category I
are all in the diagonal---see Eq.~(\ref{DG})---and they may,
therefore,
be absorbed by redefining the diagonal matrices $\Gamma_k$
as renormalized coupling matrices.
This concludes our argument,
valid at least at the one-loop level,
that in the asymptotic limit $m_R \to \infty$
one obtains the Lagrangian of Eq.~(\ref{YY})
and our model approaches a multi-scalar-doublet SM
with suppressed off-diagonal couplings in $\tilde \Gamma_j$. 

Several remarks are in order.
First,
we stress that the contributions to $z_L^\ell$ from $S^0$ exchange,
and those to $z_L^\nu$ from $S^\pm$ exchange,
are  flavor-diagonal and Hermitian and,
therefore,
they do not have anti-Hermitian components
which would interfere with the arguments presented above.
This must be so because these contributions belong to Category III. 
Second,
though in the limit of infinite right-handed scale
we were able to show---taking into account a rotation of $\nu_L$---that
the one-loop contributions of the right-handed neutrino singlets
are the same for couplings of the types $\bar\ell_R \ell_L S^0$
and $\bar\ell_R \tilde\nu_L S^-$,
this does not happen with contributions of Category III,
due to the different mass effects of charged leptons
and of massless neutrinos;
thus,
the {\em fully\/} one-loop-corrected coupling matrices $\Gamma_k$
do receive different finite parts in the 
$\bar\ell_R \ell_L S^0$ and $\bar\ell_R \tilde\nu_L S^-$ couplings
because of contributions of Category III,
an effect which is to be expected in the multi-scalar-doublet SM.
Of course,
the infinite corrections to couplings of both types are the same,
which allows for a consistent renormalization procedure.
Third,
because of the cancellation in Eq.~(\ref{cancel}),
the infinities at the vertex $V_W$
stemming from scalar corrections of Category I also cancel.
This is necessary for consistency,
since it would be impossible to absorb those infinities
into the gauge coupling constant.

\section{Decay rates}
\label{sec:decayrates}

Experimental bounds on lepton-flavor-changing processes
are found in Ref.~\cite{groom}.
For the $\tau$ the bounds on the branching ratios are of order $10^{-6}$,
but for $\mu$ decays they are five to six orders of magnitude better.
Of course,
all our formulas are easily adapted to muon decays.

In this section we shall always neglect
the masses of all final-state fermions.
Pursuing the philosophy laid out in Section~\ref{sec:orders},
we shall also assume that Yukawa couplings are of order
$Y \sim m_D/m_F$,
cf.\ Eq.~(\ref{orderY}).

First we consider the process $\tau^- \to \mu^- \gamma$.
Its matrix element may be written
\begin{equation}\label{Agamma}
\mathcal{A} \left( \tau^- \to \mu^- \gamma \right)
= e\, \varepsilon_\rho^\ast\,
\bar u_\mu i\sigma^{\rho \lambda} q_\lambda 
\left( A^\gamma_L \gamma_L + A^\gamma_R \gamma_R \right) u_\tau\, ,
\end{equation}
where $\varepsilon$ is the polarization vector of the photon.
Then the decay rate is
\begin{equation}\label{Ggamma}
\Gamma \left( \tau^- \to \mu^- \gamma \right) = \Gamma_\tau\,
\frac{48 \pi^3 \alpha}{G_F^2 m_\tau^2}
\left( \left| A^\gamma_L \right|^2 + \left| A^\gamma_R \right|^2
\right),
\end{equation}
where $\alpha$ is the electromagnetic fine-structure constant,
$G_F$ is the Fermi constant,
and
\begin{equation}
\Gamma_\tau \equiv \Gamma \left( \tau^- \to e^- \bar\nu_e \nu_\tau \right)
= \frac{G_F^2 m_\tau^5}{192 \pi^3}\, .
\end{equation}
The amplitudes in Eq.~(\ref{Agamma}) are given in our model
by Eqs.~(\ref{alphaLi}) and (\ref{alphaRi}):
\begin{equation}
A^\gamma_{L,R} = \sum_i \alpha_{L,Ri} \left( q^2 = 0 \right).
\end{equation}
In Subsection~\ref{mxlsr}
we have made for those amplitudes the order-of-magnitude estimate
$A^\gamma_{L,R} \sim Y^2 m_D / (16\pi^2 m_R^2)$.
Thus,
the branching ratio should be of order $\alpha Y^4 /(\pi G_F^2 m_R^4)$,
where we have used $m_\tau \sim m_D$.
With $G_F \approx 10^{-5}$ GeV$^{-2}$ and $m_R \sim 10^{10}$ GeV,
we see that the branching ratio would be $\sim 10^{-32}$
even if we allowed $Y$ to be of order 1.

Next we consider the lepton-flavor-changing $Z$ decay $Z \to \tau^+ \mu^-$.
Its decay amplitude may be written
\begin{equation}
\mathcal{A} \left( Z \to \tau^+ \mu^- \right)
= \frac{i g}{2 c_w}\, \varepsilon^\rho\, \bar u_\mu \gamma_\rho
\left( A^Z_L \gamma_L + A^Z_R \gamma_R \right) v_\tau\, ,
\end{equation}
where $\varepsilon$ now denotes the $Z$ polarization vector.
The decay rate is then
\begin{equation}
\Gamma \left( Z \to \tau^+ \mu^- \right)
= \frac{g^2 m_Z}{96\pi c_w^2} 
\left( \left| A^Z_L \right|^2 + \left| A^Z_R \right|^2 \right).
\end{equation}
The amplitudes $A^Z_{L,R}$ can be read off
from $\sum_{\iota=1}^4 M^\rho_{Z \iota}$,
computed in Section~\ref{sec:Z}.
As they are of order $m_D^2 /m_R^2$,
once again one finds a ridiculously small decay rate.

We now discuss neutral-scalar decay,
which is {\em not\/} suppressed by inverse powers of $m_R$
whenever $n_H \geq 2$.
The matrix element is
\begin{equation}
\mathcal{A} \left( S_b^0 \to \tau^+ \mu^- \right)
= \bar u_\mu \left( A^b_L \gamma_L + A^b_R \gamma_R \right) v_\tau\, .
\end{equation}
Then the decay rate is
\begin{equation}
\Gamma \left( S_b^0 \to \tau^+ \mu^- \right)
= \frac{m_b}{16\pi}
\left( \left| A^b_L \right|^2 + \left| A^b_R \right|^2 \right).
\end{equation}
In our model we may identify 
\begin{equation}
A^b_L \gamma_L + A^b_R \gamma_R = \tilde{M}_b^\infty\, ,
\end{equation}
with $\tilde{M}_b^\infty$ given by Eq.~(\ref{unsup})
without the spinors.
This is not suppressed by any power of $m_R^{-1}$
but it contains three Yukawa couplings.
Thus,
the decay rate should in general be very small
due to a factor $Y^6 / (16 \pi^2)^2$.
In any case this is not very interesting,
since no fundamental neutral scalar as been observed up to now.

Finally we consider our model process $\tau^- \to \mu^- e^+ e^-$.
Rather general formulas for the decay rate can
be found in Refs.~\cite{ilakovac,okada}.
Comparing our result of Section~\ref{sec:gamma}
for the photon sub-process with the results in Ref.~\cite{ilakovac},
we conclude that this contribution has a branching ratio of order
$[\alpha^2/ (G_F^2 m_R^4)] (m_D/m_F)^4 / 16\pi^2$.
In the $Z$ sub-process (Section~\ref{sec:Z}), 
$W$ exchange dominates over the exchange of charged scalars.
Thus we estimate for the branching ratio of this sub-process 
$[\alpha_w^4/(G_F^2 m_R^4)] (m_D/m_Z)^4$ 
(of course,
$m_Z \sim m_F$),
where $\alpha_w = g^2/(4\pi)$ is the weak fine-structure constant.
A similar suppression is found for the box sub-process
of Section~\ref{sec:box}. 
Therefore we may safely neglect all those contributions,
including interference terms,
and concentrate only on the neutral-scalar sub-process
computed in Section~\ref{sec:S}.
The amplitude is then written
\begin{equation}
\mathcal{A} \left( \tau^- \to \mu^- e^+ e^- \right) = \sum_b 
\bar u_\mu \left( \lambda_b \gamma_L + \rho_b \gamma_R \right) u_\tau\,
\bar u_e \left( \Gamma_b \gamma_L + \Gamma_b^\ast \gamma_R \right) v_e\, ,
\label{hxjsl}
\end{equation}
which is modeled according to Eq.~(\ref{barMb}).
Thus,
we are making the identifications $\Gamma_b = (\Gamma_b)_{ee}$ and,
assuming more than one scalar doublet,
\begin{equation}
\bar u_\mu \left( \lambda_b \gamma_L + \rho_b \gamma_R \right) u_\tau =
- \frac{M_b^\infty}{\sqrt{2}\, m_b^2}\, ,
\label{jxmaw}
\end{equation}
where we have used the approximation $q^2 \ll m_b^2$.
From Eq.~(\ref{hxjsl}) one obtains
\be
\frac{\Gamma \left( \tau^- \to \mu^- e^+ e^- \right)}{\Gamma_\tau} = 
\frac{1}{32\, G_F^2} \left( 
\left| \sum_b \lambda_b \Gamma_b \right|^2 +
\left| \sum_b \lambda_b \Gamma_b^\ast \right|^2 +
\left| \sum_b \rho_b \Gamma_b \right|^2 +
\left| \sum_b \rho_b \Gamma_b^\ast \right|^2
\right).
\ee
Both $\rho_b$ and $\lambda_b$ are of the form $Y^3 / (16 \pi^2 m_H^2)$,
where $m_H$ is a typical neutral-scalar mass,
which should be of order $m_F$.
Therefore,
the order of magnitude of the part of the branching ratio
which is not suppressed by powers of $m_R$
is $Y^8 / (16 \pi^2 G_F m_F^2)^2$.
With $16 \pi^2 G_F m_F^2 \sim 10$ and $Y \sim 10^{-2}$ as reasonable 
numerical values, 
this branching ratio is quite small, of order $10^{-18}$, 
yet it is much larger than the photon,
$Z$,
and box contributions,
which are all suppressed by $m_R^{-4}$.
Due to the dependence on $Y^8$, we can easily
achieve a branching ratio much larger than $10^{-18}$ by moderately
increasing $Y$.
However, in
this case it is certainly not possible to allow for $Y \sim 1$.
In particular, if we apply the present estimate to the branching ratio of 
$\mu^- \to e^- e^+ e^-$, we rather find $Y^8 \lesssim 10^{-10}$.

\section{Conclusions} \label{sec:conclusions} 

In this paper we have computed the amplitude of the
lepton-flavor-violating decay
$\tau^- \to \mu^- e^+ e^-$ in the context of the seesaw model 
with an arbitrary number $n_H$ of Higgs doublets, 
but with the assumption that the (tree-level)
Yukawa couplings conserve lepton flavor.
Our calculation is easily adapted to other lepton-flavor-violating decays
of the same type,
e.g.\ $\mu^- \to e^- e^+ e^-$;
moreover,
the parts of the calculation in which
$\tau^- \to \mu^- e^+ e^-$ proceeds via an intermediate photon
or an intermediate $Z$ boson are also applicable
to lepton-flavor-changing decays
like $\tau^\pm \to \mu^\pm \gamma$ and $Z \to \tau^\pm \mu^\mp$,
respectively.
As a function of the right-handed scale $m_R$ we have found the following
behavior of the decay amplitudes: 
\begin{eqnarray}
\mathcal{A}(\tau^- \to \mu^- \gamma)  & \propto & 1/m_R^2 \,, \\
\mathcal{A}(Z \to \tau^+ \mu^-)       & \propto & 1/m_R^2 \,, \\[1mm]
\mathcal{A}(\tau^- \to \mu^- e^+ e^-) & \propto & 
\left\{ \begin{array}{ll}
1/m_R^2 & \mathrm{for} \quad n_H = 1 \,, \\
\mathrm{constant} & \mathrm{for} \quad n_H > 1 \,.
\end{array}
\right.
\end{eqnarray}
The partial amplitudes for $\tau^- \to \mu^- e^+ e^-$ 
which behave like $m_R^{-2}$ are, 
in general,
suppressed by at least a factor $m_D m_F / m_R^2 \sim 10^{-18}$;
here we have assumed that
the elements of the Dirac neutrino mass matrix $M_D$
are of order $m_D \sim 1$ GeV,
the Fermi scale is of order $m_F \sim 100$ GeV,
and $m_R \sim 10^ {10}$ GeV.\footnote{We want to stress that our
philosophy is different from the one employed,
e.g.\ in Refs.~\cite{ilakovac,cvetic}.
We assume that $m_D$ is of the order of $m_\mu$ or $m_\tau$,
thus $m_R$ is very large in order to implement the seesaw mechanism.
In Refs.~\cite{ilakovac,cvetic} it is assumed that $m_R \sim 1$ TeV,
or smaller,
in order to obtain effects in lepton-flavor-violating processes.}
The exception to this suppression occurs
precisely when there is more than one scalar doublet,
for then there are contributions to $\tau^- \to \mu^- {S^0_b}^\ast$, 
where $S^0_b$ is a neutral scalar,
which are not suppressed by inverse powers of $m_R$.
These unsuppressed contributions originate from exchange 
of the charged scalars $S_a^ \pm$;
the corresponding Feynman diagrams are shown in Fig.~\ref{unsuppressed}.
We have found that
the vertices $\tau^- \to \mu^- \gamma^\ast$
and $\tau^- \to \mu^- Z^\ast$ (and therefore also
the decay amplitudes for $\tau^- \to \mu^- \gamma$ and
$Z \to \tau^+ \mu^-$),
which had already been computed before by other authors
in the case of only one scalar doublet,
are, even in the case of many doublets,
suppressed by $m_D^2 / m_R^2 \sim 10^{-20}$. 
Furthermore, also the box diagrams behave like $1/m_R^2$ for an arbitrary
$n_H$.\footnote{Note that the processes $\tau^- \to \mu^- \mu^- e^+$
and $\tau^- \to e^- e^- \mu^+$ can,
at one-loop level,
proceed only via box diagrams \cite{ilakovac}.}
Therefore, with the above numerical assumptions,
all classes of diagrams, except those with intermediate neutral scalars in the
case of more than one Higgs doublet, 
are basically irrelevant for the process~(\ref{process}).

The unsuppressed contribution to the vertex $\tau^- \to \mu^- {S^0_b}^\ast$
is found in Eq.~(\ref{unsup}),
with the quantities $A_L$ and $A_R$ given by Eqs.~(\ref{AL}) and (\ref{AR}),
respectively.
That unsuppressed contribution
has three Yukawa couplings and a denominator $16 \pi^2$;
if we assume each Yukawa coupling
to be of order $m_D / m_F \sim 10^{-2}$,
we obtain an overall order of magnitude $10^{-8}$.
This is certainly very small,
but anyway much larger than anything which is suppressed by $m_R^{-2}$.
Moreover,
we may allow for Yukawa couplings larger than $10^{-2}$.

Estimating the branching ratio for the non-suppressed decay
$\tau^- \to \mu^- e^+ e^-$, we obtain \cite{GL01}
$Y^8 / \left( 16 \pi^2 G_F m_H^2 \right)^2 \sim 10^{-18}$,
where we have taken $m_H \sim m_F$ for a typical neutral-scalar mass,
$Y \sim m_D/m_F$ for a typical Yukawa coupling,
and with the same assumptions as before for the scales $m_D$,
$m_F$,
and $m_R$.
This branching ratio,
valid for $n_H > 1$,
is not suppressed by the large scale $m_R$,
but it is suppressed instead by the eighth power
of a typical Yukawa coupling $Y$.
We want to stress, however,
that the estimate $10^{-18}$ for the branching ratio
of $\tau^- \to \mu^- e^+ e^-$ is rather crude;
having in mind the remarks at the end of the previous paragraph,
it could easily be several orders of magnitude larger
due to its dependence on $Y^8$.
On the other hand,
the branching ratio for $\tau^- \to \mu^- \gamma$ is suppressed by
$\alpha Y^4 / (\pi G_F^2 m_R^4) \sim 10^{-40}$.

We have studied in detail the non-decoupling in the Higgs sector
for $m_R \to \infty$.
In this limit our model approaches
a multi-Higgs-doublet Standard Model
with lepton-flavor non-diagonal Yukawa couplings,
the off-diagonal couplings being,
however,
suppressed.
The situation here reminds somehow
the Minimal Supersymmetric Standard Model,
which has two Higgs doublets,
where non-decoupling in the Higgs sector has been found
when the SUSY scale is made much larger than the Fermi scale
but the masses of all scalars are kept of order $m_F$
\cite{herrero}.

The model discussed here with lepton-flavor-diagonal Yukawa couplings
was put forward in Ref.~\cite{GL01} as a framework 
for imposing large or maximal neutrino mixing.
Here we have shown in a detailed way that,
despite the soft breaking of the lepton numbers
at the very large scale $m_R$,
the branching ratios of lepton-flavor-violating processes
remain very small.
We have identified the class of processes
whose vertices are not suppressed by $m_R^{-2}$. In this class the
most promising example for future experiments
is $\mu^- \to e^- e^+ e^-$; this
non-suppression requires more than one Higgs doublet.
Those vertices suppressed by $m_R^{-2}$ lead to branching ratios
far beyond present
or future
experimental limits. 
Thus we have a viable model with the interesting feature
that neutrino mixing has its origin
at the ultra-high scale $m_R$---the order of magnitude of the masses of the
right-handed neutrino singlets, which is at the same time,
via the seesaw mechanism,
responsible for the smallness of the light-neutrino masses.

\section*{Acknowledgements}

W.G. thanks A. Bartl,
W. Bernreuther,
H. Neufeld,
and A. Santamar\'\i a for helpful discussions.
He is particularly indebted to G. Ecker for clarifying remarks
on the renormalization of fermion self-energies with mixing.

\newpage

\begin{appendix}

\section{Formalism for the scalar sector}
\label{app:formalism}

\subsection{$S_a^\pm$ and $S_b^0$}
\label{sub:Sab}

We allow for an arbitrary number $n_H$ of scalar doublets $\phi_k$
($k=1,2, \ldots, n_H$)
and use the notation
\begin{equation}\label{VEV}
\phi_k = \left( \begin{array}{c}
\varphi_k^+ \\ \varphi_k^0 \end{array} \right),
\quad \mathrm{with} \quad
\left\langle 0 \left| \varphi_k^0 \right| 0 \right\rangle
= \frac{v_k}{\sqrt{2}}\, .
\end{equation}
We then write
\begin{equation}
\varphi^0_k = \frac{v_k}{\sqrt{2}} + \phik\, ,
\quad \mbox{hence} \quad
\left\langle 0 \left| \phik \right| 0 \right\rangle = 0\, .
\end{equation}
The quadratic terms in the scalar potential are written
\begin{equation}
V_\mathrm{mass} = \sum_{i,j} \left[
\varphi^-_i \left( \mathcal{M}^2_+ \right)_{ij}\, \varphi^+_j
+ A_{ij}\, \mbox{Re}\, \phii\, \mbox{Re}\, \phij
+ B_{ij}\, \mbox{Im}\, \phii\, \mbox{Im}\, \phij
+ 2 C_{ij}\, \mbox{Re}\, \phii\, \mbox{Im}\, \phij
\right].
\label{ABC}
\end{equation}
The matrix $\mathcal{M}^2_+$ is complex and Hermitian,
while the matrices $A$ and $B$ are real and symmetric;
$C$ is real but otherwise arbitrary.
They are all $n_H \times n_H$ matrices.
The eigenvalue equations are
\begin{eqnarray}
\label{eigena}
\mathcal{M}^2_+\, a &=& m_a^2\, a\, ,
\\
\label{eigenb}
\mathcal{M}^2_0
\left( \begin{array}{c} \mbox{Re}\, b \\ \mbox{Im}\, b \end{array}
\right)
&=&
\left( \begin{array}{cc} A & C \\ C^T & B \end{array} \right)
\left( \begin{array}{c} \mbox{Re}\, b \\ \mbox{Im}\, b \end{array} \right)
= m_b^2
\left( \begin{array}{c} \mbox{Re}\, b \\ \mbox{Im}\, b \end{array} \right),
\end{eqnarray}
where $a$ and $b = \mbox{Re}\, b + i\, \mbox{Im}\, b$
are $n_H \times 1$ complex vectors.
The orthonormality equations for the eigenvectors are
\begin{equation}
\sum_k a_k^* a'_k = a^\dagger a' = \delta_{a a'}\, ,
\quad
\sum_k \left( \mbox{Re}\, b_k\, \mbox{Re}\, b'_k +
              \mbox{Im}\, b_k\, \mbox{Im}\, b'_k \right) =
\mbox{Re}\, (b^\dagger b') = \delta_{b b'}\, ,
\label{nalqo}
\end{equation}
\begin{equation}
\sum_a a^\ast_k a_{k^\prime}
= \sum_b \mbox{Re}\, b_k\, \mbox{Re}\, b_{k^\prime}
= \sum_b \mbox{Im}\, b_k\, \mbox{Im}\, b_{k^\prime}
= \delta_{k k^\prime}\, ,
\quad \mbox{and} \quad
\sum_b \mbox{Re}\, b_k\, \mbox{Im}\, b_{k^\prime} = 0\, .
\end{equation}
The physical (mass-eigenstate) charged scalars $S_a^+$
and the physical neutral scalars $S_b^0$ are given by \cite{grimus89}
\begin{equation}
S_a^+ = \sum_k a_k^* \varphi^+_k \quad \mbox{and} \quad
S_b^0 = \sqrt{2}\, \sum_k \mbox{Re} \left( b_k^* \phik \right),
\end{equation}
respectively;
or,
equivalently,
\begin{equation}\label{phiS}
\phi_k =
\left( \begin{array}{c}
\sum_a a_k S_a^+ \\
\left( v_k + \sum_b b_k S_b^0 \right) / \sqrt{2}
\end{array} \right).
\end{equation}
Indeed,
one then has
\begin{equation}
V_\mathrm{mass} = \sum_a m_a^2\, S_a^- S_a^+ +
\frac{1}{2} \sum_b\, m_b^2 \left( S_b^0 \right)^2.
\end{equation}

\subsection{The mass matrices of the scalars}
\label{sub:massmatrices}

The scalar potential is
\begin{equation}\label{V}
V = \sum_{i,j} \mu^2_{ij} \phi^\dagger_i \phi_j +
\sum_{i,j,k,l} \lambda_{ijkl}
\left( \phi^\dagger_i \phi_j \right)
\left( \phi^\dagger_k \phi_l \right),
\end{equation}
with
\begin{equation}\label{Vcond}
\mu^2_{ij} = \left( \mu^2_{ji} \right)^\ast,
\quad
\lambda_{ijkl} = \lambda_{klij} \, ,
\quad \mbox{and} \quad
\lambda_{ijkl} = \lambda_{jilk}^\ast\, .
\end{equation}
The mass matrix of the charged scalars is obtained
from the potential in Eq.~(\ref{V}):
\begin{equation}
\mathcal{M}^2_+ = \mu^2 + \Lambda\, ,
\quad \mbox{with} \quad
\Lambda_{ij} = \sum_{k,l} \lambda_{ijkl}\, v_k^* v_l\, .
\label{M+}
\end{equation}
The matrix $\Lambda$ is Hermitian.
We proceed to find out the terms in $V$
which are quadratic in the neutral scalars.
We need to define two more $n_H \times n_H$ matrices,
$K$ which is symmetric
(but, in general, not Hermitian)
and $K^\prime$ which is Hermitian:
\begin{equation}
K_{ik} = \sum_{j,l} \lambda_{ijkl}\, v_j v_l
\quad \mathrm{and} \quad
K^\prime_{il} = \sum_{j,k} \lambda_{ijkl}\, v_j v_k^*\, .
\label{K'}
\end{equation}
Computing the real matrices defined in Eq.~(\ref{ABC}),
we arrive at the result
\begin{eqnarray}
A &=& \mbox{Re} \left( \mu^2 + \Lambda + K^\prime \right)
+ \mbox{Re}\, K\, , \\
B &=& \mbox{Re} \left( \mu^2 + \Lambda + K^\prime \right)
- \mbox{Re}\, K\, , \\
C &=& - \mbox{Im} \left( \mu^2 + \Lambda + K^\prime \right)
- \mbox{Im}\, K\, .
\end{eqnarray}
These matrices determine the $2n_H \times 2n_H$ mass matrix
$\mathcal{M}^2_0$ of the neutral scalars.
Equation~(\ref{eigenb}) reads
\begin{equation}
\left( \mu^2 + \Lambda + K^\prime \right) b + K b^* = m_b^2 b\, .
\label{b}
\end{equation}

\subsection{The Goldstone bosons}
\label{sub:goldstone}

The Goldstone bosons corresponding to the longitudinal modes
of the $W$ and $Z$ vector bosons are given,
respectively,
by
\begin{equation}\label{aWbZ}
a_W = \frac{1}{v} \left( v_1 , v_2, \ldots, v_{n_H} \right)^T
\quad \mbox{and} \quad
b_Z = \frac{i}{v} \left( v_1 , v_2, \ldots, v_{n_H} \right)^T,
\end{equation}
where
\begin{equation}
v = \sqrt{\left| v_1 \right|^2 + \left| v_2 \right|^2
+ \cdots + \left| v_{n_H} \right|^2}
= \frac{2 m_W}{g}\, .
\end{equation}
They correspond to zero eigenvalues of the mass matrices
$\mathcal{M}^2_+$ and $\mathcal{M}^2_0$,
respectively.
Indeed,
making the replacement $\phi_k \to
\langle 0 | \phi_k | 0 \rangle
= ( 0, v_k/\sqrt{2} )^T$ in the scalar potential of Eq.~(\ref{V})
and enforcing the condition that this is a stability point of $V$,
one obtains
\begin{equation}
\frac{\partial}{\partial v_i^*}\,
V \left( \left\langle 0 \left| \phi_j \right| 0 \right\rangle \right)
= \frac{1}{2} \left( \sum_j \mu^2_{ij}\, v_j
+ \sum_{j,k,l} \lambda_{ijkl}\, v_j v_k^* v_l \right)
= \frac{1}{2}\, \sum_j \left( \mathcal{M}^2_+ \right)_{ij} v_j = 0\, .
\label{VEVcond}
\end{equation}
Thus,
$\mathcal{M}^2_+ a_W = 0$.
Furthermore,
inserting $b_Z$ into Eq.~(\ref{b}) one obtains
\begin{equation}
\left( \mu^2 + \Lambda + K^\prime \right) b_Z + K b_Z^*
= i \left( K^\prime a_W - K a_W^* \right) = 0\, ,
\end{equation}
where we have used $b_Z = i a_W$,
$\left( \mu^2 + \Lambda \right) a_W = 0$,
and the definitions of $K$ and $K^\prime$ in Eq.~(\ref{K'}).
Thus,
$m_{a_W} = 0$ and $m_{b_Z} = 0$.\footnote{After addition
of the gauge-fixing terms to the Lagrangian,
we eventually have $m_{a_W} = m_W$.
Since we use the unitary gauge for the $Z$ boson,
the neutral Goldstone boson $S^0_{b_Z}$ does not appear in our calculation.}

\subsection{Feynman rules for some gauge vertices}
\label{sub:feynman}

The covariant derivative of the scalar doublets is
(we use the notation of Ref.~\cite{book})
\begin{eqnarray}
D_\mu \phi_k &=&
\left[ \partial_\mu
- i\, \frac{g}{\sqrt{2}}
\left( \begin{array}{cc} 0 & W^+_\mu \\ W^-_\mu & 0 \end{array} \right)
- i\, \frac{g}{2 c_w} \left( \begin{array}{cc}
c_w^2 - s_w^2 & 0 \\ 0 & - 1 \end{array} \right) Z_\mu
\right. \nonumber\\ & & \left.
+ i e \left( \begin{array}{cc} A_\mu & 0 \\ 0 & 0 \end{array} \right)
\right]
\left( \begin{array}{c}
\sum_a a_k S_a^+ \\
\left( v_k + \sum_b b_k S_b^0 \right) / \sqrt{2}
\end{array} \right).
\label{covarder2}
\end{eqnarray}
The weak mixing angle $\theta_w$,
the positron charge $e$,
and the $SU(2)$ gauge coupling $g$ are related by $e = g s_w$,
where $c_w = \cos{\theta_w}$ and $s_w = \sin{\theta_w}$.

The covariant derivative in Eq.~(\ref{covarder2})
leads to $W^\pm S_a^\mp S^0_b$ couplings given by
\begin{equation}
{\cal L} = \cdots
+ i\, \frac{g}{2}\, W^+_\mu \sum_{a,b} a^\dagger b
\left( S_a^-\, \partial^\mu S_b^0  - S_b^0\, \partial^\mu S_a^- \right)
+ \mbox{H.c.}\, ,
\label{W+S-S0}
\end{equation}
where the sums over $a$ and $b$ include the Goldstone bosons.
The Feynman rules for the vertices $W^+_\mu S_a^- S^0_b$
and $W^-_\mu S_a^+ S^0_b$ are,
therefore,
\begin{equation}
i\, \frac{g}{2}\, a^\dagger b \left( p_b - p_a \right)^\mu
\quad \mbox{and}\ \quad
i\, \frac{g}{2}\, b^\dagger a \left( p_a - p_b \right)^\mu,
\label{3-vertex}
\end{equation}
respectively.
In Eq.~(\ref{3-vertex}),
$p_b$ and $p_a$ are the incoming momenta of $S_b^0$ and of $S_a^\pm$,
respectively.

>From the covariant derivative (\ref{covarder2})
one also derives the coupling
\be
{\cal L} = \cdots + g m_W W^+_\mu {W^-}^\mu \sum_{b \neq b_Z} 
\left( a_W^\dagger b \right) S_b^0 \,.
\ee
Note that
$a_W^\dagger b$ is real for all $b \neq b_Z$,
because of the second Eq.~(\ref{nalqo})
and  $b_Z = i a_W$ (see Subsection \ref{sub:goldstone}).

The covariant derivative in Eq.~(\ref{covarder2})
also yields the $W^\pm S_a^\mp \gamma$
and $W^\pm S_a^\mp Z$ couplings,
given by
\be
{\cal L} = \cdots
- m_W \left( e A^\mu + \frac{g s_w^2}{c_w}\, Z^\mu \right)
\left( W^+_\mu G^- + W^-_\mu G^+ \right),
\label{W+G-A}
\ee
where $G^\pm \equiv S_{a_W}^\pm$ are the charged Goldstone bosons.
We emphasize that the charged Goldstone bosons are the only charged scalars
which have a vertex with $W^\pm \gamma$ and with $W^\pm Z$.

We also need the couplings of the photon and the $Z$ boson to the
charged scalars:
\be
{\cal L} = \cdots + i \left(- e A^\mu + g\, \frac{c_w^2-s_w^2}{2c_w}\,
Z^\mu \right) \sum_a 
\left( S_a^- \partial_\mu S_a^+ - S_a^+ \partial_\mu S_a^- \right) \,.
\ee
Here the sum includes the charged Goldstone boson.

For the three-gauge-boson vertices
$W^\pm W^\mp \gamma$ and $W^\pm W^\mp Z$ see, 
for instance,
Ref.~\cite{book}. 

\subsection{Feynman rules for some scalar vertices}
\label{sub:scalar}

Let us consider the contributions to ${\cal L} = \cdots - V$
with one positively-charged scalar,
one negatively-charged scalar,
and neutral scalars:
\begin{equation}
{\cal L} = \cdots - \sum_{i,j,k,l}\, \sum_{a,a'\!,b,b'}
\lambda_{ijkl}\, a_i^* S_a^- a'_j S_{a'}^+
\left( v_k^* + b_k^* S_b^0 \right) \left( v_l + b'_l S_{b'}^0 \right).
\end{equation}
We pick from this expression the vertex $S_a^- S_{a^\prime}^+ S_b^0$:
\be
{\cal L} = \cdots - \sum_{a,a^\prime \! ,b} C_{aa'b}
S_a^- S_{a^\prime}^+ S_b^0\, ,
\quad \mathrm{with} \quad
C_{aa^\prime b} = C_{a^\prime ab}^\ast =
\sum_{i,j,k,l} \lambda_{ijkl} a_i^\ast a^\prime_j
\left( v_k^\ast b_l + v_l b_k^\ast \right).
\label{jxhsl}
\ee
In the particular case $a^\prime = a_W$,
with $a_W$ given by Eq.~(\ref{aWbZ}),
we obtain
\begin{equation}
{\cal L} = \cdots
- \frac{g}{2m_W} \sum_{a,b} \sum_{i,j,k,l} a_i^* v_j \lambda_{ijkl}
\left( v_k^* b_l + b_k^* v_l \right) G^+ S_a^- S_b^0
+ \mbox{H.c.}
\end{equation}
With the definitions of Eq.~(\ref{K'}) and using Eq.~(\ref{b}),
we perform the simplification
\begin{equation}
\sum_{i,j,k,l} a_i^* v_j \lambda_{ijkl} \left( v_k^* b_l + b_k^* v_l \right)
= \sum_i a_i^* \left( K^\prime b + K b^* \right)_i
= \left( m_b^2 - m_a^2 \right) a^\dagger b\, .
\end{equation}
We thus arrive at the interaction
\begin{equation}
{\cal L} = \cdots 
+ \frac{g}{2 m_W}\, \sum_{a,b}
a^\dagger b \left( m_a^2 - m_b^2 \right) G^+ S_a^- S_b^0
+ \mbox{H.c.}\, ,
\label{uskzy}
\end{equation}
which is the counterpart of Eq.~(\ref{W+S-S0}) in $R_\xi$ gauges.
Therefore,
we have
\begin{equation}\label{C12}
C_{aa_Wb} = \frac{g}{2 m_W} \left( m_b^2 - m_a^2 \right) a^\dagger b
\quad \mathrm{and} \quad 
C_{a_Wab} = \frac{g}{2 m_W} \left( m_b^2 - m_a^2 \right) b^\dagger a\, ,
\end{equation}
and 
the Feynman rules for the vertices $G^+ S_a^- S_b^0$
and $G^- S_a^+ S_b^0$ are
\begin{equation}
i\, \frac{g}{2m_W}\, a^\dagger b \left( m_a^2 - m_b^2 \right)
\quad \mbox{and}\ \quad
i\, \frac{g}{2m_W}\, b^\dagger a \left( m_a^2 - m_b^2 \right),
\label{3S-vertex}
\end{equation}
respectively.

In Eq.~(\ref{uskzy}) we may consider the particular case $a = a_W$.
Remember that
$a_W^\dagger b$ is real for all $b \neq b_Z$.
Using $m_{a_W} = 0$ 
(see Subsection~\ref{sub:goldstone}),
we obtain 
\begin{equation}\label{C3}
C_{a_Wa_Wb} = \frac{g}{2 m_W}\, m_b^2 \left( a_W^\dagger b \right) \,,
\end{equation}
and the corresponding Feynman rule for the vertex $G^+ G^- S^0_b$:
\be
- i\, \frac{g}{2 m_W}\, m_b^2 \left( a_W^\dagger b \right).
\ee

\section{The fermion self-energy and its renormalization}
\label{app:renormalization}

\subsection{On-shell renormalization conditions}
\label{sub:onshell}

Most of the material in this Subsection can be found, e.g.\ in
Refs.~\cite{soares,kniehl}.
We include it in order to make our paper self-contained,
since the renormalization procedure outlined here
is used in Section~\ref{sec:asymptotic}. 

Let us consider a theory with $n$ Dirac fermions,
e.g.\ charged leptons.
We use a matrix notation whenever possible.
The unit matrix is not distinguished from the number 1.
The one-loop fermion self-energy is
\begin{equation}
- i \Sigma (p) = - i \left\{ p \hskip-5pt / \left[
A_L \left( p^2 \right) \gamma_L + A_R \left( p^2 \right) \gamma_R \right] + 
B_L \left( p^2 \right) \gamma_L + B_R \left( p^2 \right) \gamma_R \right\}.
\end{equation}
The quantities $A_{L,R}$ and $B_{L,R}$ are  $n \times n$ matrices.
They constitute the result of the one-loop calculation.
We assume that there are no absorptive parts
in the one-loop self-energy diagrams, then
\begin{equation}\label{AB}
\left( A_L \right)_{\gamma\beta}^\ast
= \left( A_L \right)_{\beta\gamma} \,, \quad
\left( A_R \right)_{\gamma\beta}^\ast
= \left( A_R \right)_{\beta\gamma} \,, \quad \mathrm{and} \quad
\left( B_L \right)_{\gamma\beta}^\ast
= \left( B_R \right)_{\beta\gamma}\, . 
\end{equation}
We shall also assume that the fermion masses are non-degenerate.

We want to study the one-loop renormalization of the fermion fields.
The bare chiral fields are denoted $\ell_{\alpha L}$ and $\ell_{\alpha R}$,
and the renormalized ones $\ell^r_{\alpha L}$ and $\ell^r_{\alpha R}$:
\begin{equation}
\ell_{L,R} = \sqrt{Z_{L,R}}\, \ell^r_{L,R} = 
\left( 1 + \frac{1}{2}\, z_{L,R} \right) \ell^r_{L,R} \,.
\end{equation}
The $z_{L,R}$ are  $n \times n$ matrices.
The diagonal matrix of the bare fermion masses $m_\alpha$
is $M_\ell = \mbox{diag}\, (m_1, \ldots, m_n)$.
We write $M_\ell = M_\ell^r + \delta M_\ell$,
where $M_\ell^r$ is the diagonal matrix of the renormalized masses,
and $\delta M_\ell = \mbox{diag}\, (\delta m_1, \ldots, \delta m_n)$.
Then,
the \emph{renormalized} fermion self-energy is
\begin{eqnarray}
-i \Sigma^r (p) &=& -i \Sigma (p)
+
i\, \frac{1}{2} \left( z_L + z_L^\dagger \right) p \hskip-5pt / \gamma_L + 
i\, \frac{1}{2} \left( z_R + z_R^\dagger \right) p \hskip-5pt / \gamma_R 
\nonumber \\ && -
i \left( \delta M_\ell + \frac{1}{2}\, M_\ell z_L + 
\frac{1}{2}\, z_R^\dagger M_\ell \right) \gamma_L 
- i \left( \delta M_\ell + \frac{1}{2}\, z_L^\dagger M_\ell + 
\frac{1}{2}\, M_\ell z_R \right) \gamma_R \,. \hspace*{8mm}
\end{eqnarray}
The terms in $\Sigma^r (p)$ beyond those in $\Sigma (p)$
originate in the counter-terms for the masses and for the wave functions.

Up to now the renormalized self-energy has no precise meaning.
Now we fix the meaning of this notion by requiring that
$\Sigma^r$ fulfills the following conditions:\footnote{Condition 1 is
stated in Ref.~\cite{soares}. Both conditions are stated 
in Ref.~\cite{kniehl}.}
\begin{eqnarray}\label{cond1}
\mbox{Condition 1:} & \hphantom{xxx} &
\Sigma^r_{\beta \alpha} (p)
\mbox{\raisebox{-5pt}{$\Big|$} %
$\hskip-7pt$ \raisebox{-7pt}{$\scriptstyle p^2 = m_\alpha^2$}} \,
u_\alpha = 0\, , \quad \forall\, \alpha, \beta \,;
\\[1mm]
\label{cond2}
\mbox{Condition 2:} & \hphantom{xxx} &
\lim_{p^2 \to m_\alpha^2}
\left[ S_F (p) \right]_{\alpha\alpha}
\left( p \hskip-5pt / - m_\alpha \right)
u_\alpha = u_\alpha\, ,
\quad \forall\, \alpha \, ,
\end{eqnarray}
where the fermion propagator is given by
\begin{equation}
S_F (p) = \frac{1}{p \hskip-5pt / - M_\ell - \Sigma^r (p)} \,.
\end{equation}
We are using the notation $u_\alpha \equiv u( \vec{p}, m_\alpha, s)$,
where the four-spinor is to be taken with mass $m_\alpha$,
three-momentum $\vec p$,
and polarization $s$;
it satisfies $p \hskip-5pt / u_\alpha = m_\alpha u_\alpha$.
The above equations are the conditions for on-shell self-energy
(or propagator)
renormalization \cite{soares,kniehl,weinberg,ross,sakakibara},
and they include rotating back the renormalized fermion fields
into the physical basis.
Condition 2 fixes the residuum of the propagator $S_F (p)$
at the pole $p^2 = m_\alpha^2$ to be 1.

Exploiting first Condition 1,
i.e.\
Eq.~(\ref{cond1}),
it is easy to see that it leads to the following relations,
which originate in the coefficients of $\gamma_R$ and $\gamma_L$,
respectively:
\begin{eqnarray}
\label{R}
m_\alpha A_L \left( m_\alpha^2 \right)_{\beta\alpha}
+ B_R \left( m_\alpha^2 \right)_{\beta\alpha}
- \frac{1}{2}\, m_\alpha (z_L)_{\beta\alpha} 
+ \frac{1}{2}\, m_\beta (z_R)_{\beta\alpha} 
+ \delta m_\alpha \delta_{\beta\alpha} &=& 0\, , \\
\label{L}
m_\alpha A_R \left( m_\alpha^2 \right)_{\beta\alpha}
+ B_L \left( m_\alpha^2 \right)_{\beta\alpha}
- \frac{1}{2}\,m_\alpha (z_R)_{\beta\alpha}
+ \frac{1}{2}\, m_\beta (z_L)_{\beta\alpha} 
+ \delta m_\alpha \delta_{\beta\alpha} &=& 0 \,.
\end{eqnarray}
These equations may be solved to give \cite{soares,kniehl},
for $\alpha \neq \beta$,
\begin{eqnarray}
\frac{1}{2}\, (z_L)_{\beta\alpha} &=& \frac{
m_\alpha^2 A_L \left( m_\alpha^2 \right)_{\beta\alpha} +
m_\alpha m_\beta A_R \left( m_\alpha^2 \right)_{\beta\alpha} +
m_\beta B_L \left( m_\alpha^2 \right)_{\beta\alpha} +
m_\alpha B_R \left( m_\alpha^2 \right)_{\beta\alpha}}
{m_\alpha^2 - m_\beta^2} \, ,
\label{zL} \\
\frac{1}{2}\, (z_R)_{\beta\alpha} &=& \frac{
m_\alpha^2 A_R \left( m_\alpha^2 \right)_{\beta\alpha} +
m_\alpha m_\beta A_L \left( m_\alpha^2 \right)_{\beta\alpha} +
m_\beta B_R \left( m_\alpha^2 \right)_{\beta\alpha} +
m_\alpha B_L \left( m_\alpha^2 \right)_{\beta\alpha}}
{m_\alpha^2 - m_\beta^2}\, , 
\label{zR}
\end{eqnarray}
and, for $\alpha = \beta$,
\begin{eqnarray}
- 2\, \delta m_\alpha & = & m_\alpha \left[
A_L \left( m_\alpha^2 \right)_{\alpha\alpha}
+ A_R \left( m_\alpha^2 \right)_{\alpha\alpha}
\right] + B_L \left( m_\alpha^2 \right)_{\alpha\alpha} +
B_R \left( m_\alpha^2 \right)_{\alpha\alpha},
\label{deltam} \\
m_\alpha \left( z_L - z_R \right)_{\alpha\alpha} & = & m_\alpha \left[
A_L \left( m_\alpha^2 \right)_{\alpha\alpha}
- A_R \left( m_\alpha^2 \right)_{\alpha\alpha}
\right] - B_L \left( m_\alpha^2 \right)_{\alpha\alpha} +
B_R \left( m_\alpha^2 \right)_{\alpha\alpha}\, . \hspace*{8mm} 
\label{diff}
\end{eqnarray}
The right-hand side of Eq.~(\ref{deltam})
is real because of Eqs.~(\ref{AB}).
Equations~(\ref{zL})--(\ref{deltam}) fix the mass counterterms
and the off-diagonal wave-function counterterms.

With Eqs.~(\ref{AB}),
the $z_L$ of Eq.~(\ref{zL}) 
and the $z_R$ of Eq.~(\ref{zR}) agree with the corresponding
self-energy renormalization for $\alpha \neq \beta$ in
Ref.~\cite{soares},
after one specializes the functions $B_{L,R}$ to the forms used there.

Only the difference $(z_L - z_R)_{\alpha\alpha}$ is
determined by Eq.~(\ref{diff}).
In order to fix 
$(z_L)_{\alpha\alpha}$ and $(z_R)_{\alpha\alpha}$ separately,
one has to invoke Condition 2.
Equation~(\ref{cond2}) is equivalent to
\begin{equation}\label{cond2a}
\lim_{p^2 \to m_\alpha^2} 
\frac{1}{p \hskip-5pt / - m_\alpha}\, \Sigma^r_{\alpha\alpha}(p)\,
u_\alpha = 0\, , \quad \forall\, \alpha \,,
\end{equation}
which is better suited for the further procedure. 
Firstly,
we expand the matrix functions $A_{L,R}(p^2)$ and $B_{L,R}(p^2)$
in $\Sigma (p)$ around $p^2 = m_\alpha^2$;
for instance,
\begin{equation}
A_L \left( p^2 \right) = A_L \left( m_\alpha^2 \right)
+ \left( p^2-m_\alpha^2 \right) A_L^\prime \left( m_\alpha^2 \right)
+ \frac{1}{2} \left( p^2-m_\alpha^2 \right)^2 A_L^{\prime \prime}
\left( m_\alpha^2 \right) +
\cdots \, .
\end{equation}
Secondly, we use Eqs.~(\ref{R}) and (\ref{L}) to obtain
\begin{eqnarray}
\Sigma^r_{\alpha\alpha}(p) & = &
\left[ A_L \left( m_\alpha^2 \right)
- \mathrm{Re}\, z_L \right]_{\alpha \alpha}
\left( p \hskip-5pt / \gamma_L - m_\alpha \gamma_R \right)
\nonumber \\ & &
+
\left[ A_R \left( m_\alpha^2 \right)
- \mathrm{Re}\, z_R \right]_{\alpha \alpha}
\left( p \hskip-5pt / \gamma_R - m_\alpha \gamma_L \right)
\nonumber \\ & &
+ \left( p^2 - m_\alpha^2 \right)
\left[
A_L^\prime \left( m_\alpha^2 \right) p \hskip-5pt / \gamma_L
+
A_R^\prime \left( m_\alpha^2 \right) p \hskip-5pt / \gamma_R
\right. \nonumber \\ & & \left.
+
B_L^\prime \left( m_\alpha^2 \right) \gamma_L
+
B_R^\prime \left( m_\alpha^2 \right) \gamma_R
\right]_{\alpha \alpha}
\nonumber \\ & &
+ \frac{1}{2} \left( p^2 - m_\alpha^2 \right)^2
\left[
A_L^{\prime \prime} \left( m_\alpha^2 \right) p \hskip-5pt / \gamma_L
+
A_R^{\prime \prime} \left( m_\alpha^2 \right) p \hskip-5pt / \gamma_R
\right. \nonumber \\ & & \left.
+
B_L^{\prime \prime} \left( m_\alpha^2 \right) \gamma_L
+
B_R^{\prime \prime} \left( m_\alpha^2 \right) \gamma_R
\right]_{\alpha \alpha}
+ \cdots \,,
\label{sigma}
\end{eqnarray}
where the dots indicate higher orders in $p^2-m_\alpha^2$.
It is obvious from Eq.~(\ref{sigma}) that,
when $p^2 = m_\alpha^2$,
$\Sigma^r_{\alpha \alpha} (p)\, u_\alpha = 0$.
This agrees with Condition 1 in Eq.~(\ref{cond1}).
Condition 2,
in Eq.~(\ref{cond2a}),
is nonetheless non-trivial because of the presence of the denominator
$p \hskip-5pt / - m_\alpha$,
which also vanishes when applied to $u_\alpha$.
Thirdly,
we use
\begin{equation}
\frac{1}{p \hskip-5pt / - m_\alpha}
\left( p \hskip-5pt / \gamma_L - m_\alpha \gamma_R \right)
=
\left( \gamma_L p \hskip-5pt / + m_\alpha \gamma_R \right)
\frac{1}{p \hskip-5pt / + m_\alpha}\, ,
\end{equation}
together with the analogous relation for opposite chiralities,
to obtain
\begin{eqnarray}
\frac{1}{p \hskip-5pt / - m_\alpha}\, \Sigma^r_{\alpha\alpha}(p)
&=&
\left[ A_L \left( m_\alpha^2 \right)
- \mathrm{Re}\, z_L \right]_{\alpha \alpha}
\left( \gamma_L p \hskip-5pt / + m_\alpha \gamma_R \right)
\frac{1}{p \hskip-5pt / + m_\alpha}
\nonumber \\ & &
+
\left[ A_R \left( m_\alpha^2 \right)
- \mathrm{Re}\, z_R \right]_{\alpha \alpha}
\left( \gamma_R p \hskip-5pt / + m_\alpha \gamma_L \right)
\frac{1}{p \hskip-5pt / + m_\alpha}
\nonumber \\ & &
+
\left( p \hskip-5pt / + m_\alpha \right)
\left[
A_L^\prime \left( m_\alpha^2 \right) p \hskip-5pt / \gamma_L
+
A_R^\prime \left( m_\alpha^2 \right) p \hskip-5pt / \gamma_R
\right. \nonumber \\ & & \left.
+
B_L^\prime \left( m_\alpha^2 \right) \gamma_L
+
B_R^\prime \left( m_\alpha^2 \right) \gamma_R
\right]_{\alpha \alpha}
\nonumber \\ & &
+ \frac{p^2 - m_\alpha^2}{2}
\left( p \hskip-5pt / + m_\alpha \right)
\left[
A_L^{\prime \prime} \left( m_\alpha^2 \right) p \hskip-5pt / \gamma_L
+
A_R^{\prime \prime} \left( m_\alpha^2 \right) p \hskip-5pt / \gamma_R
\right. \nonumber \\ & & \left.
+
B_L^{\prime \prime} \left( m_\alpha^2 \right) \gamma_L
+
B_R^{\prime \prime} \left( m_\alpha^2 \right) \gamma_R
\right]_{\alpha \alpha}
+ \cdots\, .
\end{eqnarray}
Therefore,
\begin{eqnarray}
\frac{1}{p \hskip-5pt / - m_\alpha}\, \Sigma^r_{\alpha\alpha}(p)\, u_\alpha
&=& \left\{
\frac{\left[ A_L \left( m_\alpha^2 \right) + A_R \left( m_\alpha^2 \right)
- \mathrm{Re} \left( z_L + z_R \right) \right]_{\alpha \alpha}}{2}
\right. \nonumber \\ & &
+ m_\alpha^2 \left[
A_L^\prime \left( m_\alpha^2 \right) + A_R^\prime \left( m_\alpha^2 \right)
\right]_{\alpha \alpha}
+ m_\alpha \left[
B_L^\prime \left( m_\alpha^2 \right) + B_R^\prime \left( m_\alpha^2 \right)
\right]_{\alpha \alpha}
\nonumber \\ & &
\left. + \mathrm{O} \left( p^2 - m_\alpha^2 \right) \right\} u_\alpha\, .
\end{eqnarray}
Finally performing the limit $p^2 \to m_\alpha^2$ to apply Condition 2,
i.e.\
Eq.~(\ref{cond2a}),
we obtain \cite{kniehl}
\begin{eqnarray}
\label{z+}
\mathrm{Re} \left( z_L + z_R \right)_{\alpha\alpha}
&=&
\left\{
A_L \left( m_\alpha^2 \right) + A_R \left( m_\alpha^2 \right) +
2 m_\alpha^2 \left[ A_L^\prime \left( m_\alpha^2 \right)
+ A_R^\prime \left( m_\alpha^2 \right) \right]
\right. \nonumber \\ & & \left.
+
2 m_\alpha \left[ B_L^\prime \left( m_\alpha^2 \right)
+ B_R^\prime \left( m_\alpha^2 \right) \right] 
\right\}_{\alpha\alpha} \,.
\end{eqnarray}
Equations~(\ref{diff}) and (\ref{z+}) together fix
the real parts of $(z_L)_{\alpha\alpha}$ and of $(z_R)_{\alpha\alpha}$.
Concerning the imaginary parts of those quantities,
Eq.~(\ref{diff}) fixes their difference, 
$\mathrm{Im}\, (z_L-z_R)_{\alpha\alpha}$,
whereas their sum remains undetermined.
This fact reflects the freedom of redefining the fields
$\ell_{\alpha L}$ and $\ell_{\alpha R}$ by transforming them with the
same phase factor.

\subsection{The equivalence of two different procedures}
\label{sub:equivalence}

In Sections \ref{sec:gamma},
\ref{sec:Z},
and \ref{sec:S} we have considered,
respectively,
the vertices $\tau \mu \gamma$,
$\tau \mu Z$,
and $\tau \mu S^0_b$.
When computing those flavor-changing vertices
we have not invoked any renormalization procedure,
but we have considered self-energy transitions $\tau \to \mu$
in the external fermion legs.
In the following we show that,
instead of adding the self-energy transitions
in the external fermion legs to the one-loop vertex,
one may equivalently apply the on-shell renormalization prescription
to the external fermion legs.
We explicitly work out this equivalence,
for an arbitrary fermion self-energy,
in the case a scalar vertex (a Yukawa coupling).
However,
it will become clear that the nature of the vertex
is irrelevant for this equivalence,
and therefore our considerations are of general validity.
In the case of the one-loop flavor-changing $Z$ vertex in the Standard Model,
this equivalence was explicitly derived in Ref.~\cite{soares}.

Let us first see the effect of the on-shell wave-function renormalization
on the Yukawa interaction of a real scalar $S^0$,
\begin{equation}\label{LS}
\mathcal{L}_\mathrm{Y} \left( S^0 \right) =
- \left( \bar \ell_R \Gamma_0 \ell_L + 
\bar \ell_L \Gamma_0^\dagger \ell_R \right) S^0\, ,
\end{equation}
where $\Gamma_0$ is the $n \times n$ coupling matrix
of unrenormalized coupling constants. 
Defining the renormalized scalar field via 
$S^0 = \sqrt{Z_S} S^0_r = (1+z_S/2) S^0_r$,
and denoting the
renormalized coupling matrix by $\Gamma = \Gamma_0 - \delta \Gamma$,
the Lagrangian of Eq.~(\ref{LS}),
written in renormalized quantities,
is
\begin{eqnarray}
\mathcal{L}_\mathrm{Y} \left( S^0 \right) & = & - \left[ 
\bar \ell_R^r \left( \Gamma + \delta \Gamma + 
\frac{1}{2} z_S \Gamma +
\frac{1}{2} z_R^\dagger \Gamma + \frac{1}{2} \Gamma z_L \right)
\ell_L^r \right.
\nonumber \\ && + \left.
\bar \ell_L^r \left( \Gamma^\dagger + \delta \Gamma^\dagger + 
\frac{1}{2} z_S \Gamma^\dagger +
\frac{1}{2} \Gamma^\dagger z_R + \frac{1}{2} z_L^\dagger \Gamma^\dagger
\right) \ell_R^r \right] S^0_r \,.
\label{LSren}
\end{eqnarray}
The quantity $\delta \Gamma$ is determined by the 
renormalization condition for $\Gamma$. 

We now consider the part of the
vertex $\ell_\alpha \to \ell_\beta S^0$ given by the wave-function
renormalization constants $z_{L,R}$. First we consider those terms
which have the coupling matrices $\Gamma$ and $\Gamma^\dagger$ to the
left of the factors $z$; only those terms can give a fermion
self-energy on the $\ell_\alpha$ leg.
For any $\gamma \neq \alpha$, using the Lagrangian
in Eq.~(\ref{LSren}) and taking into account
Eqs.~(\ref{zL}) and (\ref{zR}), we obtain (see also Ref.~\cite{soares}) 
\begin{eqnarray}
\frac{1}{2}\, \bar u_\beta \left[ 
\Gamma_{\beta\gamma}
\left( z_L \right)_{\gamma\alpha} \gamma_L +
\left( \Gamma^\dagger \right)_{\beta\gamma}
\left( z_R \right)_{\gamma\alpha} \gamma_R
\right] u_\alpha &=&
\bar u_\beta \left[ 
\Gamma_{\beta\gamma} \gamma_L
+ \left( \Gamma^\dagger \right)_{\beta\gamma} \gamma_R
\right]
\nonumber \\ & & \times\, \frac{i}{\um - m_\gamma}
\left[ -i \Sigma (p_1)_{\gamma\alpha} \right] u_\alpha \,,
\label{SR}
\end{eqnarray}
where the four-spinors $u_\alpha$ and $u_\beta$ are to be taken at
four-momenta $p_1$ and $p_2$, respectively, with $p_1^2 = m_\alpha^2$ and
$p_2^2 = m_\beta^2$.
Equation~(\ref{SR}) holds for every 
$\gamma \neq \alpha$ and demonstrates that the $z$-terms taken into
account in the left-hand side of Eq.~(\ref{SR})
exactly reproduce the self-energy insertion
in the leg $\ell_\alpha$ of the tree-level scalar vertex.

Equation~(\ref{SR}) actually does not depend on
the assumption that we are dealing with a scalar vertex.
All operations needed in order to derive Eq.~(\ref{SR})
took place to the right of the matrices $\Gamma$ and $\Gamma^\dagger$,
and they involved only the spinor $u_\alpha$.
Thus,
our scalar vertex might be replaced by any other vertex with different
Dirac structure,
and the result would be analogous.

For the terms where $\Gamma$ and $\Gamma^\dagger$ are to the right of
the wave-function renormalization constants, 
we have to consider 
$(z_{L,R}^\dagger)_{\beta\gamma} = (z_{L,R})_{\gamma\beta}^*$, which
means that the matrices $A_{L,R}$ and $B_{L,R}$, as functions of
$p^2$, must be evaluated at $m_\beta^2$. 
This contrasts with Eq.~(\ref{SR}) where they had to
be taken at $m_\alpha^2$; in Eq.~(\ref{SR}) the fermion
self-energy appears with $p_1$.
Furthermore, we now must use Eq.~(\ref{AB}).
Having this in mind, for any $\gamma \neq \beta$ we arrive at
\begin{eqnarray}
\frac{1}{2}\, \bar u_\beta \left[ 
\left( z_R^\dagger \right)_{\beta\gamma} \Gamma_{\gamma\alpha} \gamma_L +
\left( z_L^\dagger \right)_{\beta\gamma}
\left( \Gamma^\dagger \right)_{\gamma\alpha} \gamma_R
\right] u_\alpha &=&
\bar u_\beta \left[ -i \Sigma (p_2) \right]_{\beta\gamma}
\frac{i}{\dois - m_\gamma}
\nonumber \\ & & \times
\left[ \Gamma_{\gamma\alpha} \gamma_L
+ \left( \Gamma^\dagger \right)_{\gamma\alpha} \gamma_R
\right] u_\alpha \,.
\label{SL}
\end{eqnarray}

In conclusion,
in the example of a simple scalar vertex and for
$\alpha \neq \gamma \neq \beta$ one
can see the equivalence \cite{soares} of the following two procedures for
calculating the transition amplitude 
$\ell_\alpha \to \ell_\beta {S^0}^*$:
\begin{enumerate}
\item On-shell renormalization 
of the propagator in order to calculate the
wave-function renormalization constants $z_{L,R}$,
and then taking them into account at the vertex;
\item Adding the two self-energy contributions to the vertex,
without any renormalization.
\end{enumerate}

\end{appendix}

\end{document}